\documentclass[twocolumn]{aastex631}
\usepackage{amsmath}
\usepackage{multirow}
\usepackage{booktabs}
\usepackage{ragged2e}

\newcommand{\ha}{H$\alpha$}
\newcommand{\hanii}{H$\alpha$+[\ion{N}{2}]}
\newcommand{\nii}{[\ion{N}{2}]}
\newcommand{\sfrha}{SFR$_{H\alpha}$}
\newcommand{\sfruv}{SFR$_{UV}$}
\newcommand{\p}{$^\prime$}

\newcommand{\hii}{H\,{\small II}}

\newcommand{\sdssr}{{\em r}}

\newcommand{\mips}{24$\mu$m}
\newcommand{\wise}{22$\mu$m}

\newcommand{\dfuv}{DF$_{UV}$}
\newcommand{\dfha}{DF$_{H\alpha}$}
\newcommand{\dfov}{DF$_{UV+H\alpha}$}
\newcommand{\luv}{L$_{UV}$}
\newcommand{\lha}{L$_{H\alpha}$}
\submitjournal{ApJ}

\shorttitle{H$\alpha$/FUV Ratios in Nearby Galaxies}
\shortauthors{Padave et al.}

\graphicspath{{./}{figures/}}

\begin{document}

\title{DIISC-V: Variations in H$\alpha$-to-FUV Star Formation Rate Ratios Across Star-forming Regions in Nearby Galaxies}

\author[0000-0002-3472-0490]{Mansi Padave}
\affiliation{School of Earth \& Space Exploration, Arizona State University, Tempe, AZ 85287-1404, USA}
\email{mpadave@asu.edu}

\author[0000-0002-2724-8298]{Sanchayeeta Borthakur}
\affiliation{School of Earth \& Space Exploration, Arizona State University, Tempe, AZ 85287-1404, USA}

\author[0000-0003-1268-5230]{Rolf A. Jansen}
\affiliation{School of Earth \& Space Exploration, Arizona State University, Tempe, AZ 85287-1404, USA}

\author[0000-0002-8528-7340]{David A. Thilker}
\affiliation{Department of Physics \& Astronomy, Johns Hopkins University, Baltimore, MD, 21218, USA}

\author{Jacqueline Monkiewicz}
\affiliation{School of Earth \& Space Exploration, Arizona State University, Tempe, AZ 85287-1404, USA}

\author[0000-0001-8156-6281]{Rogier A. Windhorst}
\affiliation{School of Earth \& Space Exploration, Arizona State University, Tempe, AZ 85287-1404, USA}

\begin{abstract}
We present the variations in far-ultraviolet (FUV) and \ha\ star formation rates (SFR), \sfruv\ and \sfrha, respectively, at sub-kpc scales in 11 galaxies as part of the Deciphering the Interplay between the Interstellar Medium, Stars, and the Circumgalactic medium (DIISC) survey. Using archival GALEX FUV imagery and \ha+\nii\ narrowband images obtained with the Vatican Advanced Technology Telescope, we detect a total of 1335 (FUV-selected) and 1474 (\ha-selected) regions of recent
high-mass star formation, respectively. We find the \ha-to-FUV SFR ratios tend to be lower primarily for FUV-selected regions, where SFR$_{\textrm{H}\alpha}$ generally underestimates the SFR by an average factor of 2--3, for SFR $<$ 10$^{-4}$ M$_{\odot}$\,yr$^{-1}$. In contrast, the SFRs are generally observed to be consistent for \ha-selected regions. This discrepancy arises from morphological differences between the two indicators. Extended FUV morphologies and larger areas covered by FUV-only regions, along with decreasing overlap between FUV clumps and compact \hii\ regions with $R/R_{25}$ suggest that stochastic sampling of the initial mass function may be more pronounced in the outer regions of galaxies. Our observed \ha-to-FUV SFR ratios are also consistent with stochastic star formation model predictions. 
However, using larger apertures that include diffuse FUV emission results in an offset of 1 dex between \sfrha\ and \sfruv. This suggests that the observed low \ha-to-FUV SFR ratios in galaxies are likely caused by diffuse FUV emission, which can contribute $\sim$60--90\% to the total FUV flux.
\end{abstract}


\section{Introduction} \label{sec:intro}
The H$\alpha$ nebular line and far-ultraviolet (FUV) continuum are two of the most widely used tracers of star formation in galaxies \citep{kenn98, kenn12}. The FUV continuum is produced by non-ionizing UV photons from O-stars as well as B- and A-stars and probes timescales of $\sim$100 Myr. 
In contrast, H$\alpha$ emission is a byproduct of gas ionized by UV photons emanating from the photospheres of short-lived massive O-type and early B-type stars, tracing a shorter star formation timescale of $\lesssim$10 Myr. Despite their power to probe recent and current high-mass star formation, respectively, these indicators have long been scrutinized due to the discrepancy between the star formation rates (SFR) of galaxies inferred from them.

Following the standard calibrations from \cite{kenn98b}, consistent values for the SFR, invariant with time and environment, are expected from the two indicators. Nonetheless, several observations have revealed a systematic decline in the H$\alpha$-to-FUV ratios with decreasing luminosities and at low levels of star formation in galaxies \citep{buat87, bell01, igle04, meur09, lee09, boselli09}. 
It is found that while the average ratio remains constant at SFR $\gtrsim0.1$ M$_\odot$ yr$^{-1}$, it varies by a factor of 2 to an order of magnitude for SFR~$\lesssim10^{-3}$ M$_\odot$ yr$^{-1}$, also observed for low-mass dwarf galaxies \citep{lee09, lee16}.  These observations highlight the nuances in the production of H$\alpha$ and FUV photons, due to deviations from the assumptions that are foundational to SFR prescriptions: a constant star formation history (SFH) over $\sim$1 Gyr, a well-populated initial mass function (IMF) for solar metallicity, and no loss of Lyman continuum photons or dust extinction. 

Many independent studies have analyzed the variations in \ha-to-FUV ratios in galaxies to explore the deviations from the model assumptions. Non-constant or bursty SFH have been found to have a considerable impact on the ratios \citep{igle04, weisz12, emami19, kauff21}. 
Lower \ha\ emission in galaxies, especially for the low mass systems ($M<10^8 M_{\odot}$), can be explained by a lull in star formation, during which the \ha\ luminosities can drop quickly compared to the FUV luminosities \citep{mehta23}. 
Moreover, stochastic sampling of stellar and cluster IMFs means lower-mass systems are likely to have fewer high-mass clusters and, consequently, fewer high-mass stars \citep{fuma11, koda12}. 
Additionally, the escape of ionizing photons from HII regions lowers photoionization rates, and excluding the diffuse component of H$\alpha$ emission in the warm ionized medium can lead to underestimating H$\alpha$ luminosities \citep{hoopes01, eldridge11, hunter11, relano12, lee16, watkins24}. Given the abundance of evidence supporting these factors, they all likely play a role in the observed variations in \ha-to-FUV ratios.

On spatially-resolved scales, the \ha-to-FUV ratios in galaxies provide interesting insights into star formation in galactic outskirts. In search of a star formation threshold, \cite{mart01} identified truncated \ha\ radial profiles in nearby spiral galaxies. However, such sharp declines were not observed in the UV profiles, and many nearby galaxies showed UV knots at large galactocentric distances or emitted faint UV, termed Type-I and Type-II extended ultraviolet (XUV) disks \citep{thilk07}. These differences in the radial profiles have been associated both with the relative numbers of the detected UV and \ha\ knots as well as the declining fluxes in the outer disks of galaxies \citep{zaritsky07, godd10}. 
\hii\ regions have also been discovered well beyond the optical edge (R$_{25}$) of the galaxy disks  \citep{hodge69, ferg98, vanzee98, leli00, ryanweber04, werk10}, sometimes as far out as 4R$_{25}$, as is the case with M83 \citep{thilk05}. 

So far, \ha-to-FUV ratios in galaxies have been extensively investigated with integrated luminosity and SFR measurements.
On spatially-resolved scales, the ratios have been explored mainly for XUV disk galaxies or individual galaxies \citep{barn11, byun21, padave21}. As part of the Deciphering the Interplay between Star Formation, Interstellar Medium, and the Circumgalactic Medium (DIISC) survey, \cite{padave21} investigated \ha\ emission for FUV-selected stellar associations in NGC~3344. They found a considerable scatter in the \ha-to-FUV SFR ratios in the XUV region, with \sfruv$\gtrsim$\sfrha\ for the stellar associations. 
In this paper, we delve deeper into the variations between the FUV and \ha\ SFR indicators on sub-kpc scales for a larger sample and include star-forming galaxies showing signatures of stellar disk growth \citep[see][and \S\ref{sec:data} for details]{padave24} from the DIISC survey. Our primary goal in this work is to examine the morphology and spatial distribution of FUV and \ha\ emission and investigate aperture effects on the connection between \sfruv\ and \sfrha\ at sub-kpc scales. The structure of the paper is as follows. In \S\ref{sec:data}, we present our sample, observations, and data reduction and calibration procedures. The source detection on the \ha\ and FUV maps is described in \S\ref{sec:reg}. We present our results on the \ha-to-FUV SFR ratios of the detected sources and their spatial overlaps in \S\ref{sec:res}, followed by a discussion on the relation between \sfrha\ and \sfruv. Finally, we present our conclusions in \S\ref{sec:conc}.

\section{Sample \& Data} \label{sec:data}
We carried out a follow-up \ha-DIISC program to obtain \hanii\ imaging of the 34 galaxies from the DIISC survey, intending to map any \hii-regions in the outskirts at a signal-to-noise ratio (SNR) $\gtrsim5$.   
Deep narrow-band \hanii\ and broad-band Sloan-\sdssr\ (henceforth denoted \sdssr) imaging observations of the galaxies were carried out with the 1.8 m {\em Vatican Advanced Technology Telescope} (VATT) operated by the Mt. Graham International Observatory in Arizona.  
The observations took place over a total of 19 nights between 2019 March to 2021 March. 
The typical seeing for these observations was between 0\farcs8 and 2\farcs2, with a pixel scale of 0\farcs375/pixel.  
A set of seven different H$\alpha$ filters of $\sim70$~\AA\ FWHM, covering the redshifted H$\alpha$ emission lines within a range of 650--700~nm, were used for these observations. 
The total integration time ranges between 60~min and 100~min in the narrow-band \ha\ filters and 20~min in \sdssr, the latter used for continuum subtraction. The continuum-subtraction procedure is described in Section \ref{sec:hared}.

\begin{figure}[!tb]
  \centering
\includegraphics[trim = 0cm 0cm 0cm 0.cm, clip,scale=0.35]{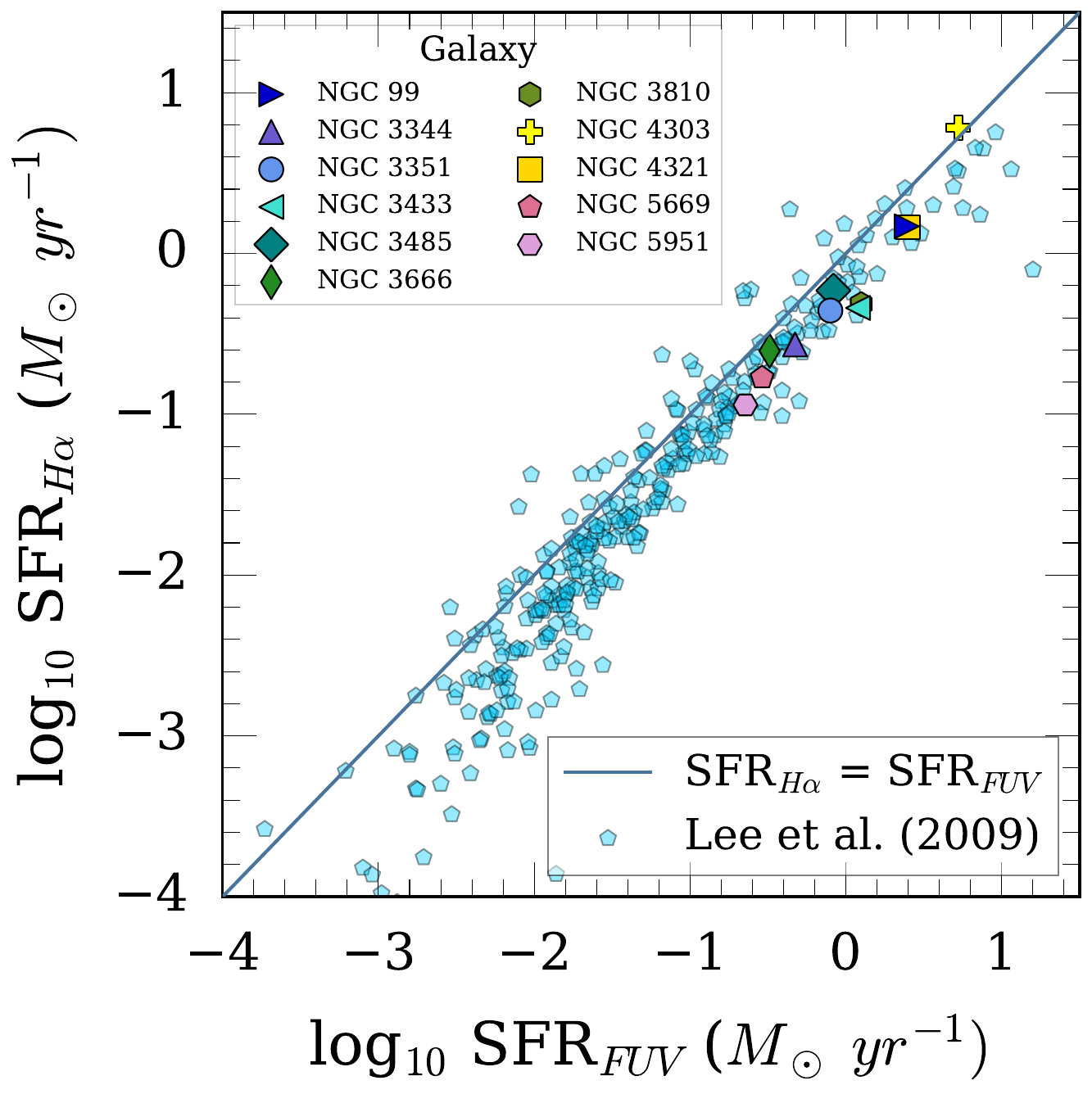}
\caption{Comparison of \sfruv\ and \sfrha\ of the sample plotted over spiral and dwarf galaxies from 11HUGS investigated in \cite{lee09}. 
The solid line represents the one-to-one correspondence between the SFRs. 
Most of our galaxies show \sfruv$>$\sfrha\ with \ha-to-FUV SFR ratios ranging from 0.38--1.15.} \label{fig:glob_sfr}
\end{figure}

To effectively investigate the \ha-to-FUV SFR ratios at sub-kpc scales, our selection criteria prioritized spatial resolution and the number of detected \hii\ regions in the \hanii\ images. After considering factors such as distance, angular diameter, availability of deep FUV data from the Galaxy Evolution Explorer \citep[GALEX;][]{mart05, morr07}, and the distribution of \hii\ regions across the disk, we chose 11 galaxies out of the 34 in the DIISC sample for the present study.
Salient features of these galaxies are noted in Table \ref{tab:gal}. Figure \ref{fig:glob_sfr} compares the \sfruv\ and \sfrha\ of our sample with the $\sim$300 spirals and dwarf galaxies from the 11~Mpc \ha\ and Ultraviolet Galaxy Survey (11HUGS) investigated by \cite{lee09}.

\begin{figure*}
  \centering
\includegraphics[trim = 0cm 0cm 1.5cm 1cm, clip,scale=0.19]{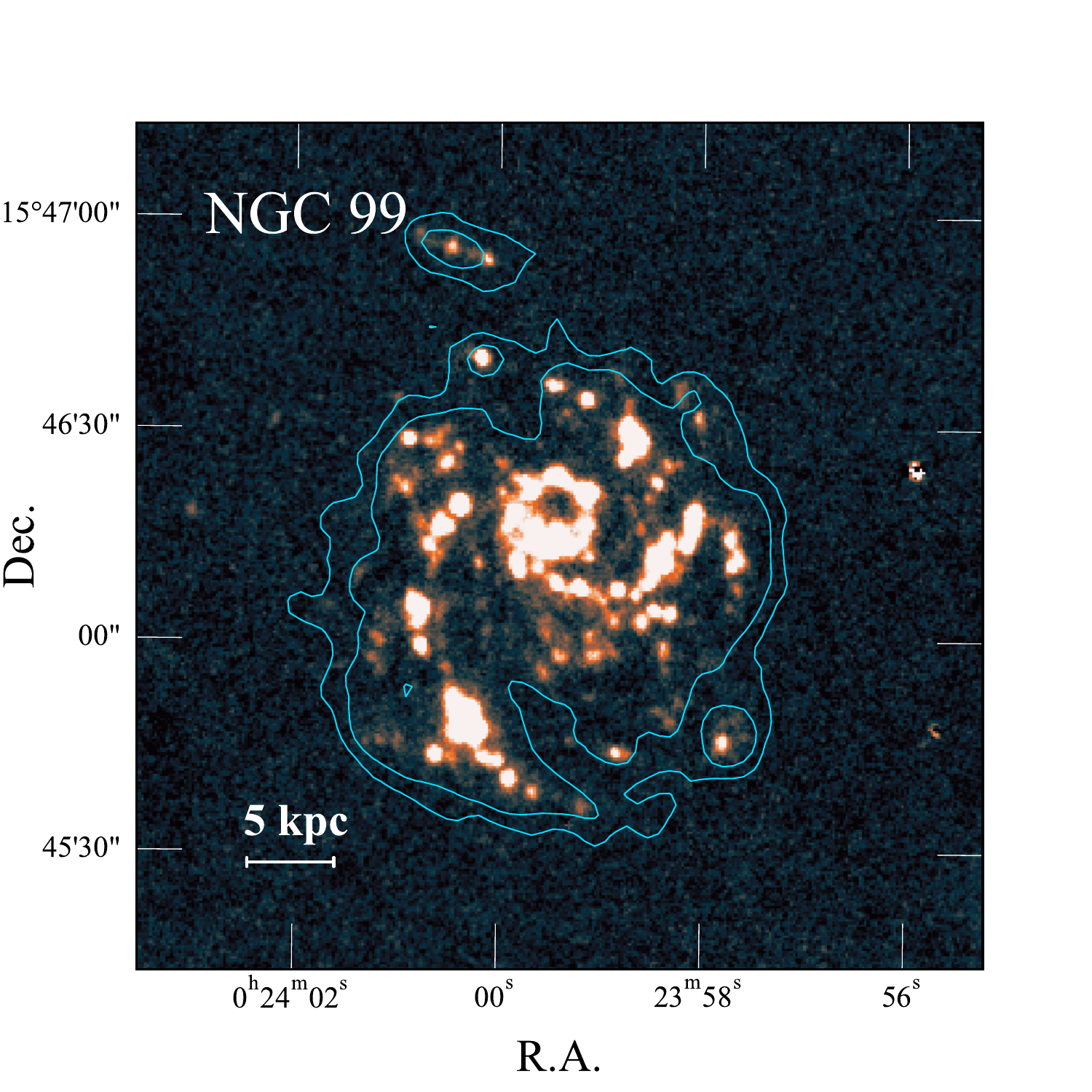}
\hspace{-0.5cm}
\includegraphics[trim = 0cm 0cm 1cm 1cm, clip,scale=0.19]{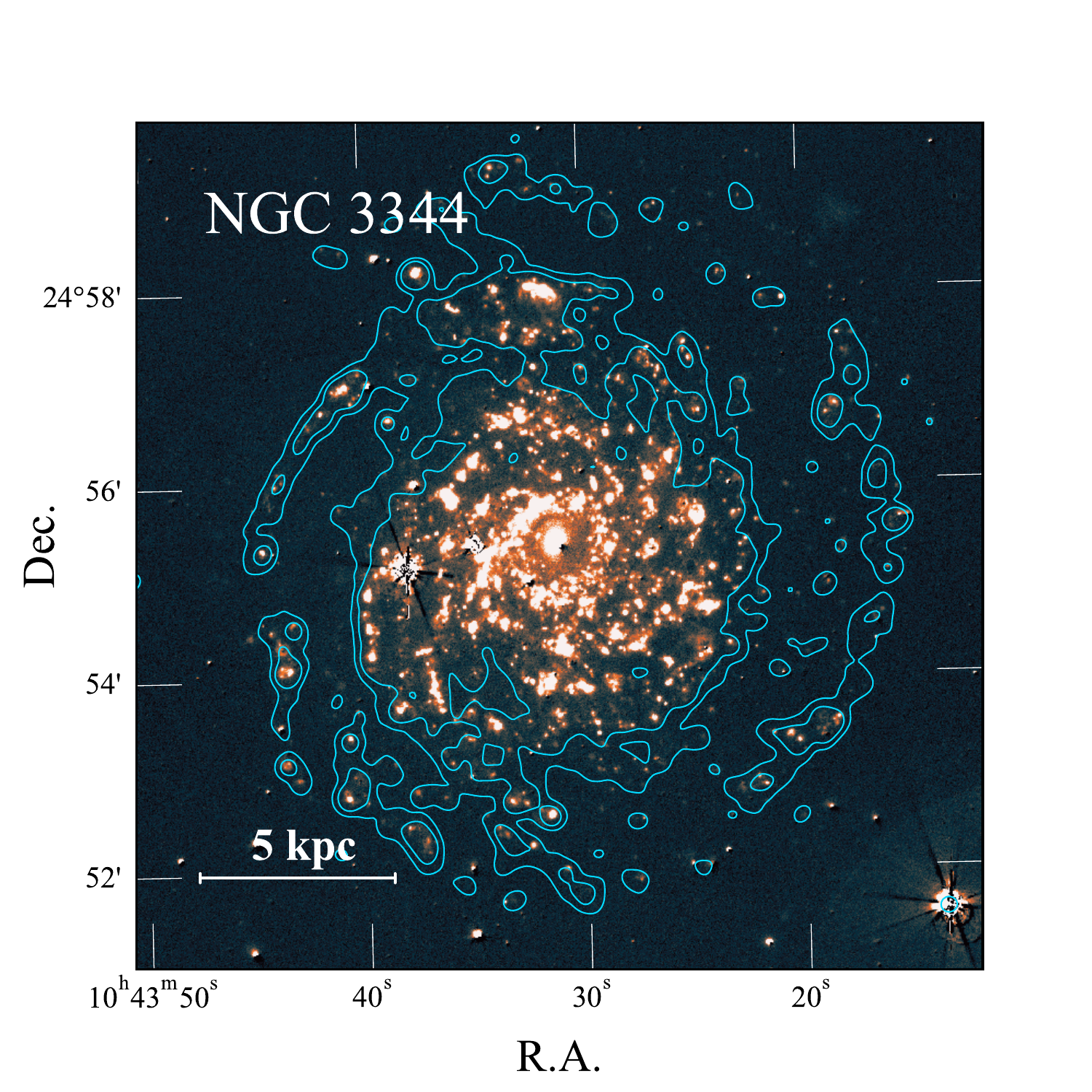}\hspace{-0.5cm}
\includegraphics[trim = 0cm 0cm 1cm 1cm, clip,scale=0.19]{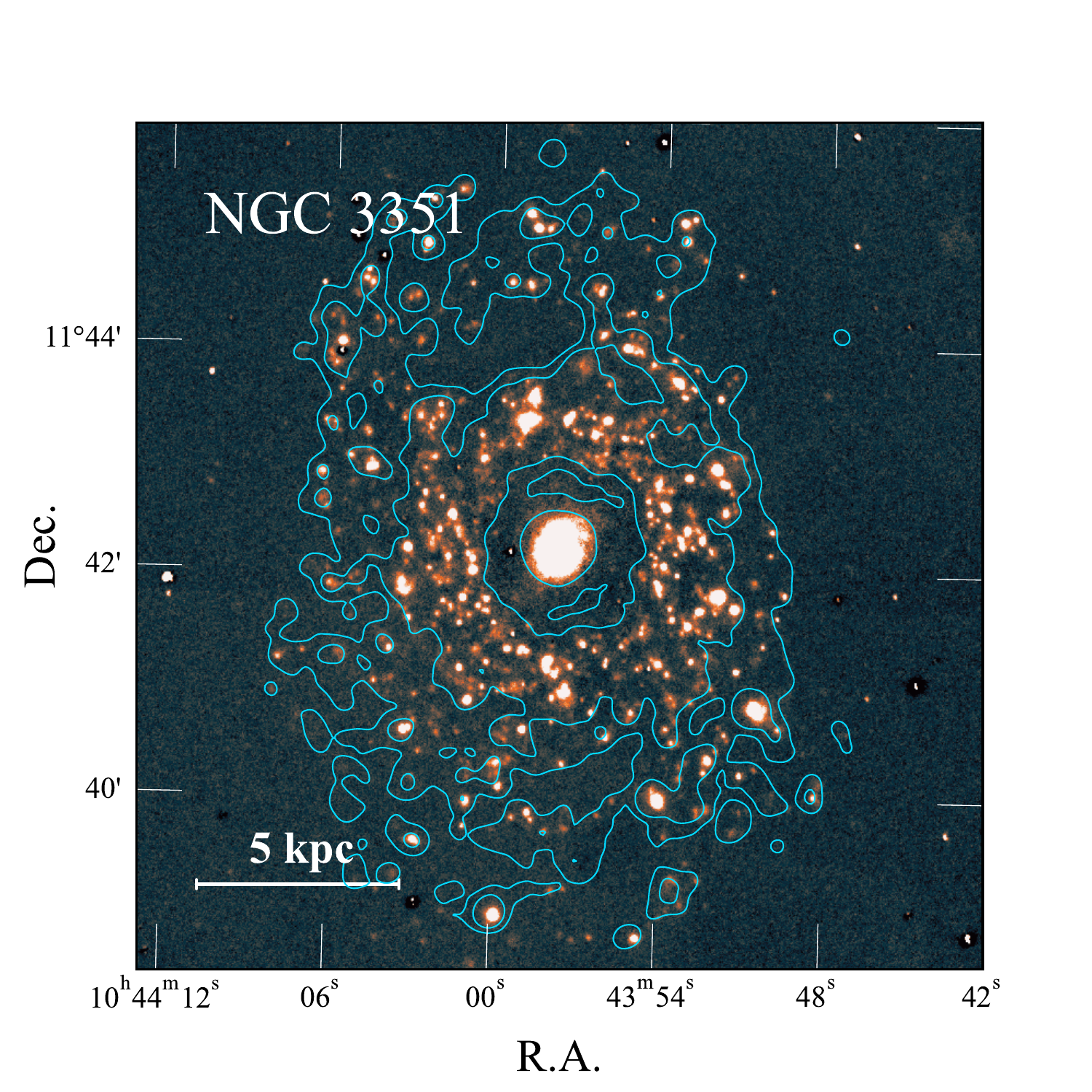}\hspace{-0.5cm}
\includegraphics[trim = 0cm 0cm 1cm 1cm, clip,scale=0.19]{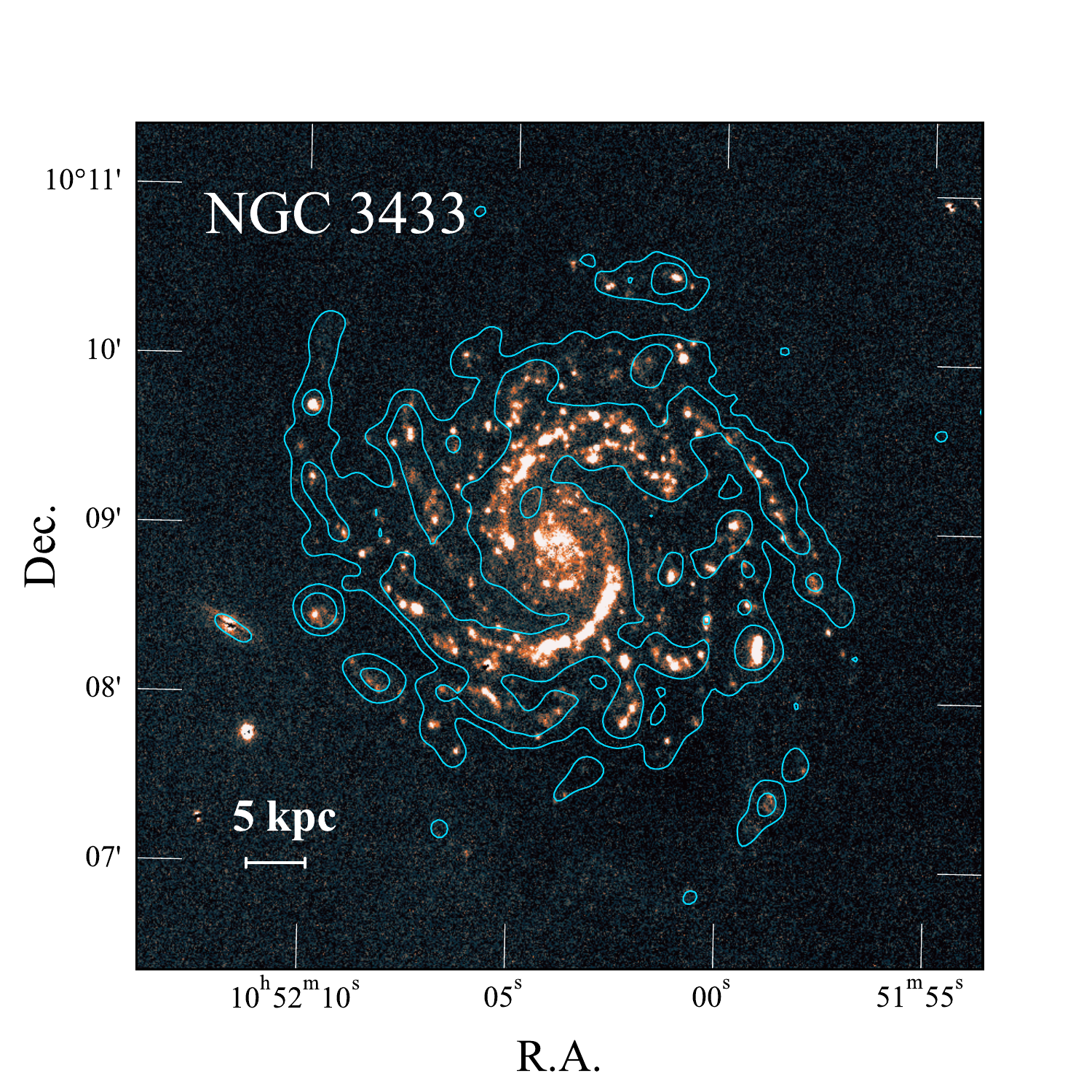}\hspace{-0.5cm}
\includegraphics[trim = 0cm 0cm 1cm 1cm, clip,scale=0.19]{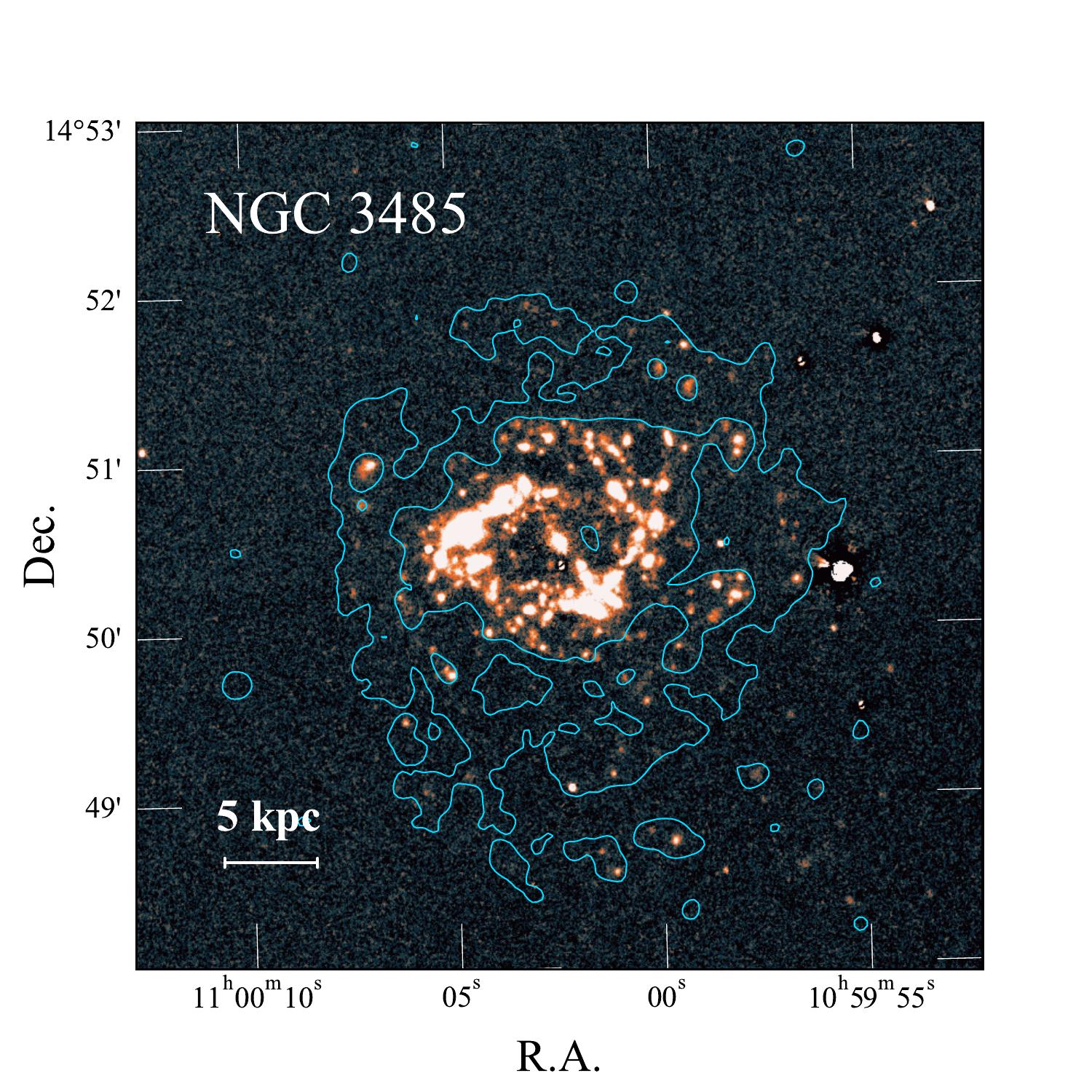}\hspace{-0.5cm}
\includegraphics[trim = 0cm 0cm 1cm 1cm, clip,scale=0.19]{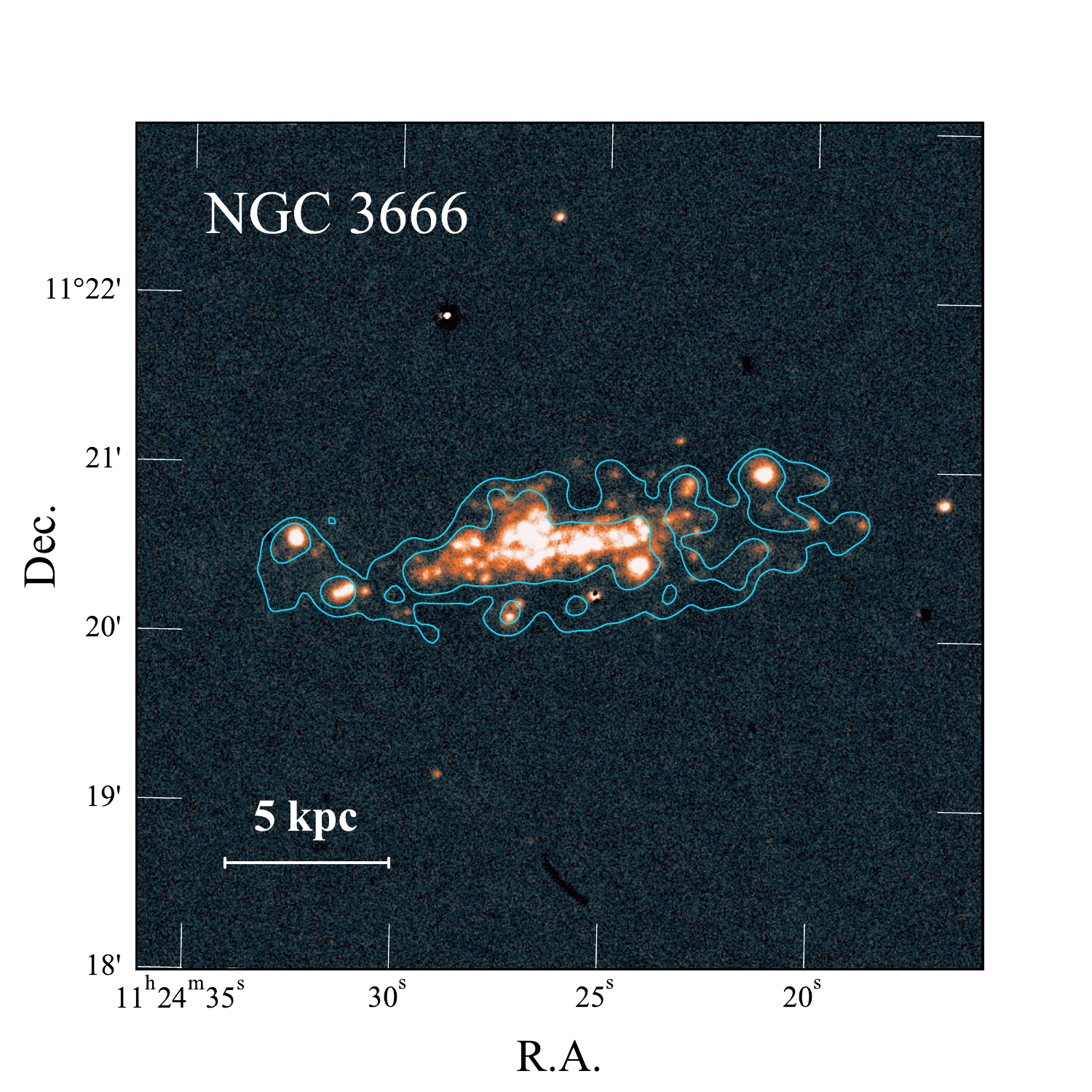}\hspace{-0.5cm}
\includegraphics[trim = 0cm 0cm 1cm 1cm, clip,scale=0.19]{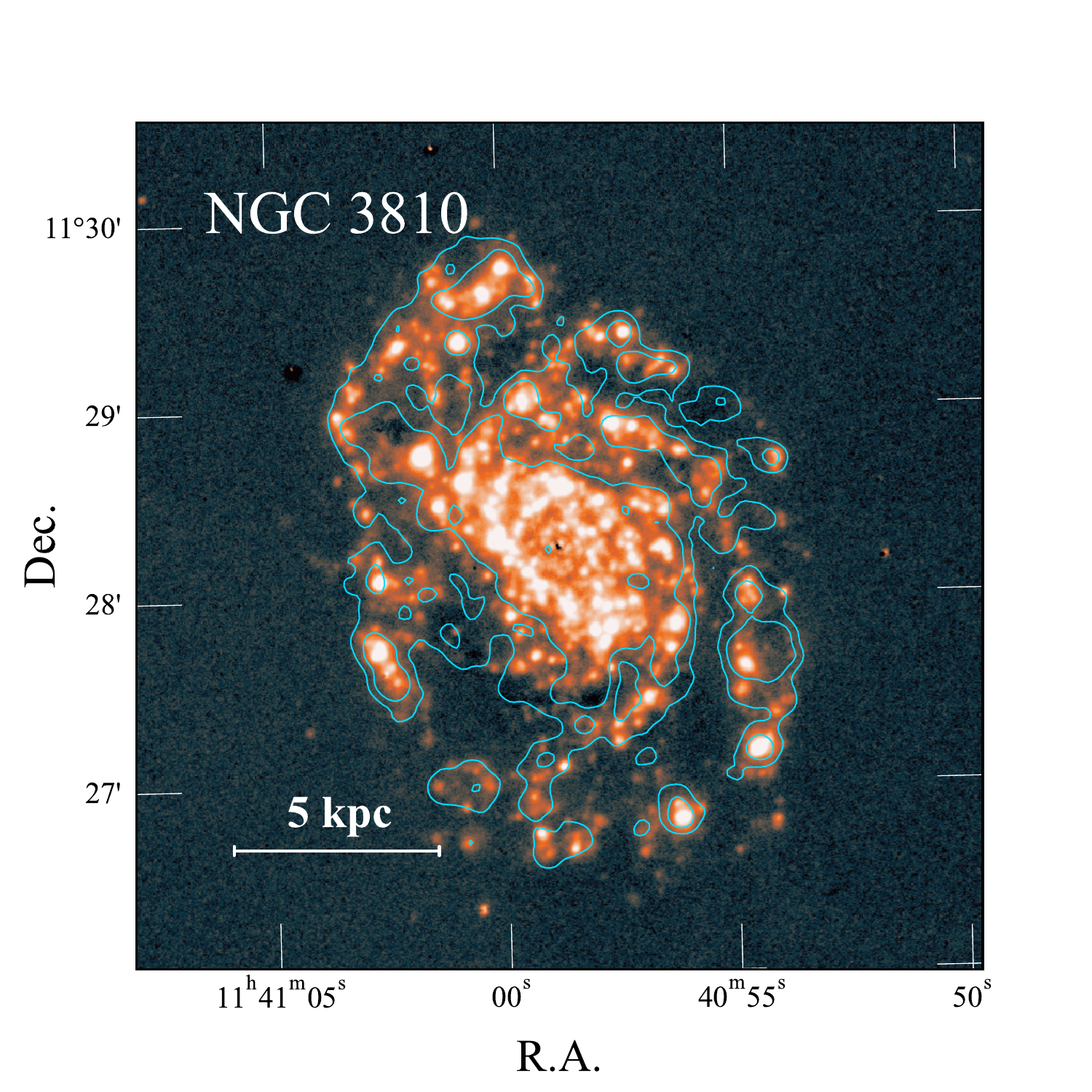}\hspace{-0.5cm}
\includegraphics[trim = 0cm 0cm 1cm 1cm, clip,scale=0.19]{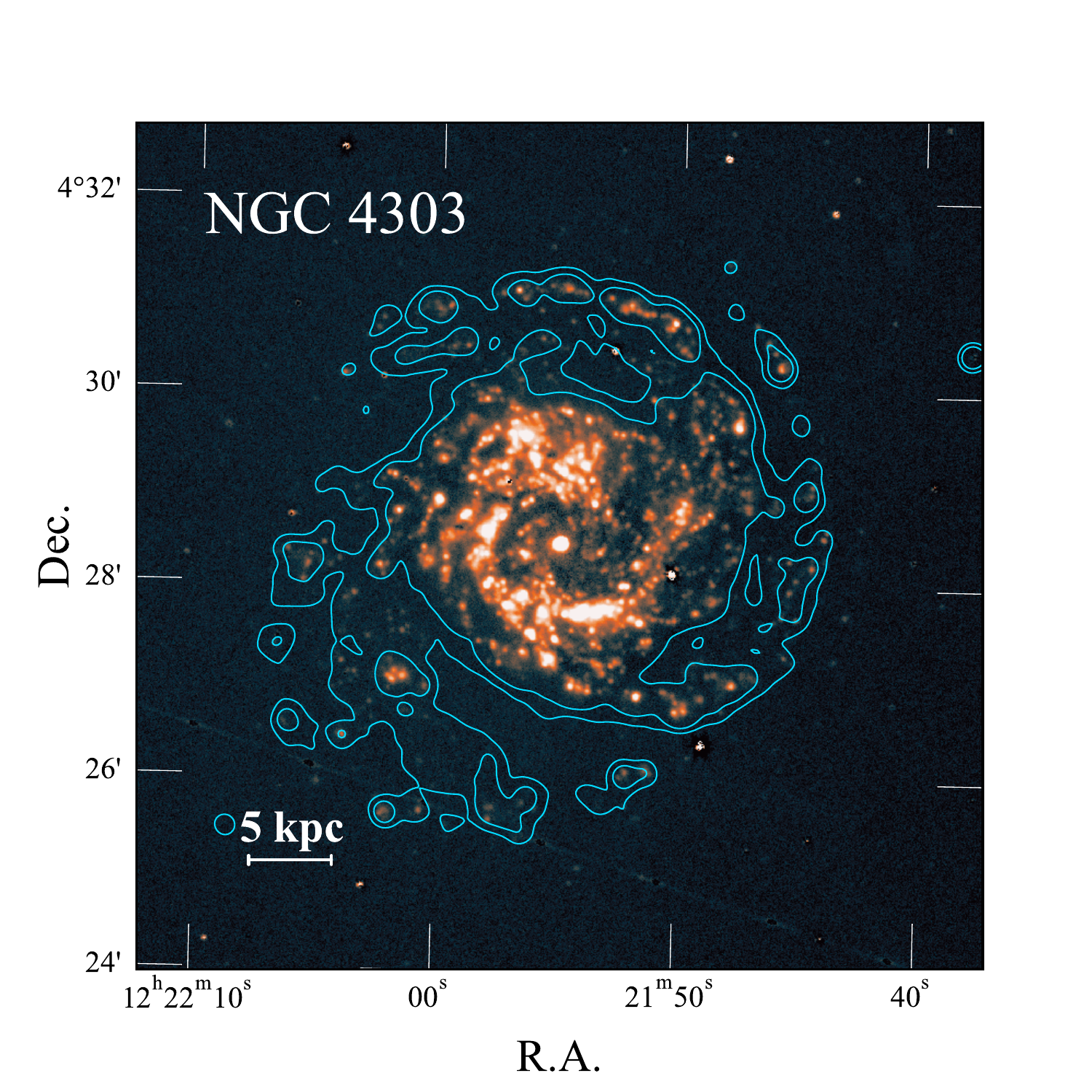}\hspace{-0.5cm}
\includegraphics[trim = 0cm 0cm 1cm 1cm, clip,scale=0.19]{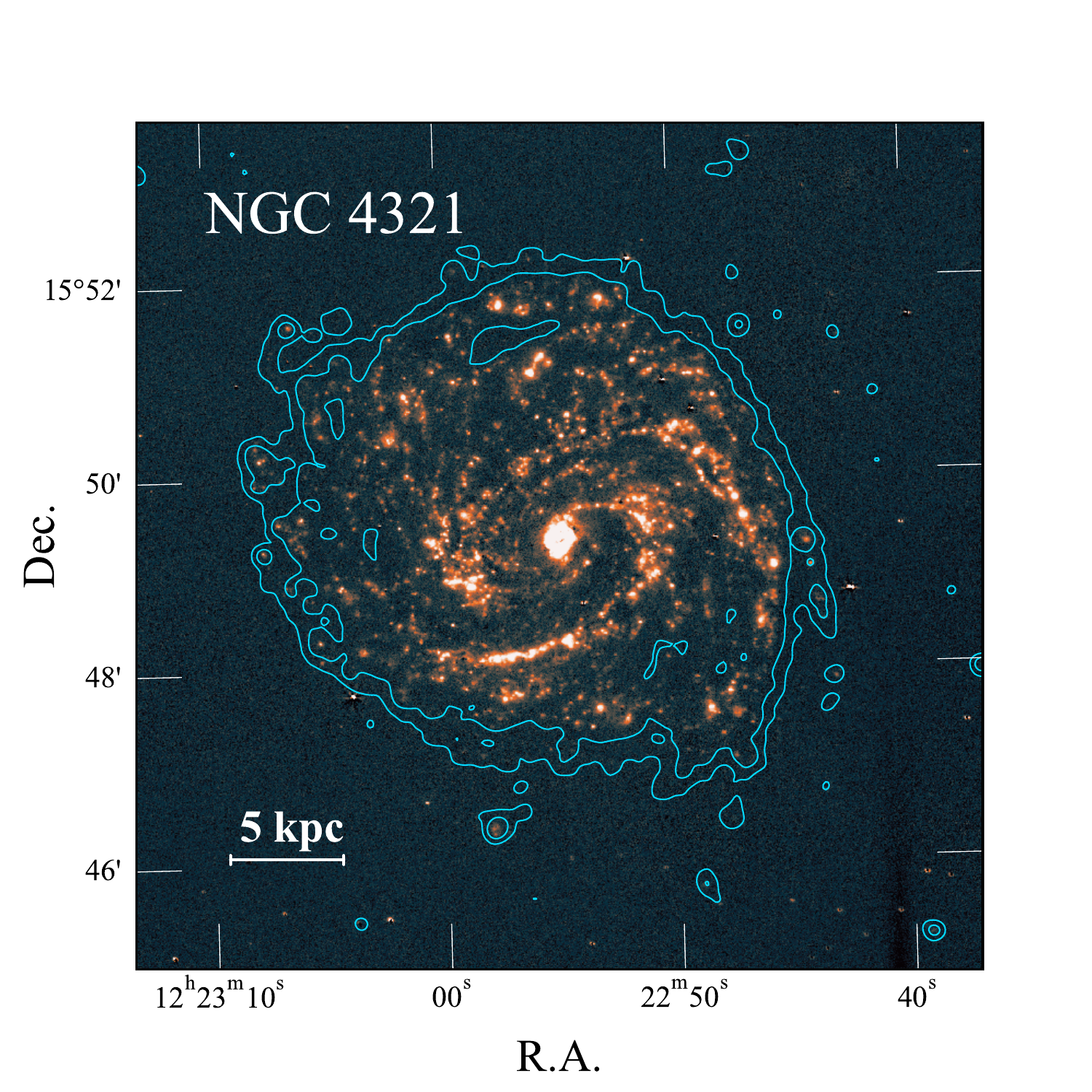}\hspace{-0.5cm}
\includegraphics[trim = 0cm 0cm 1cm 1cm, clip,scale=0.19]{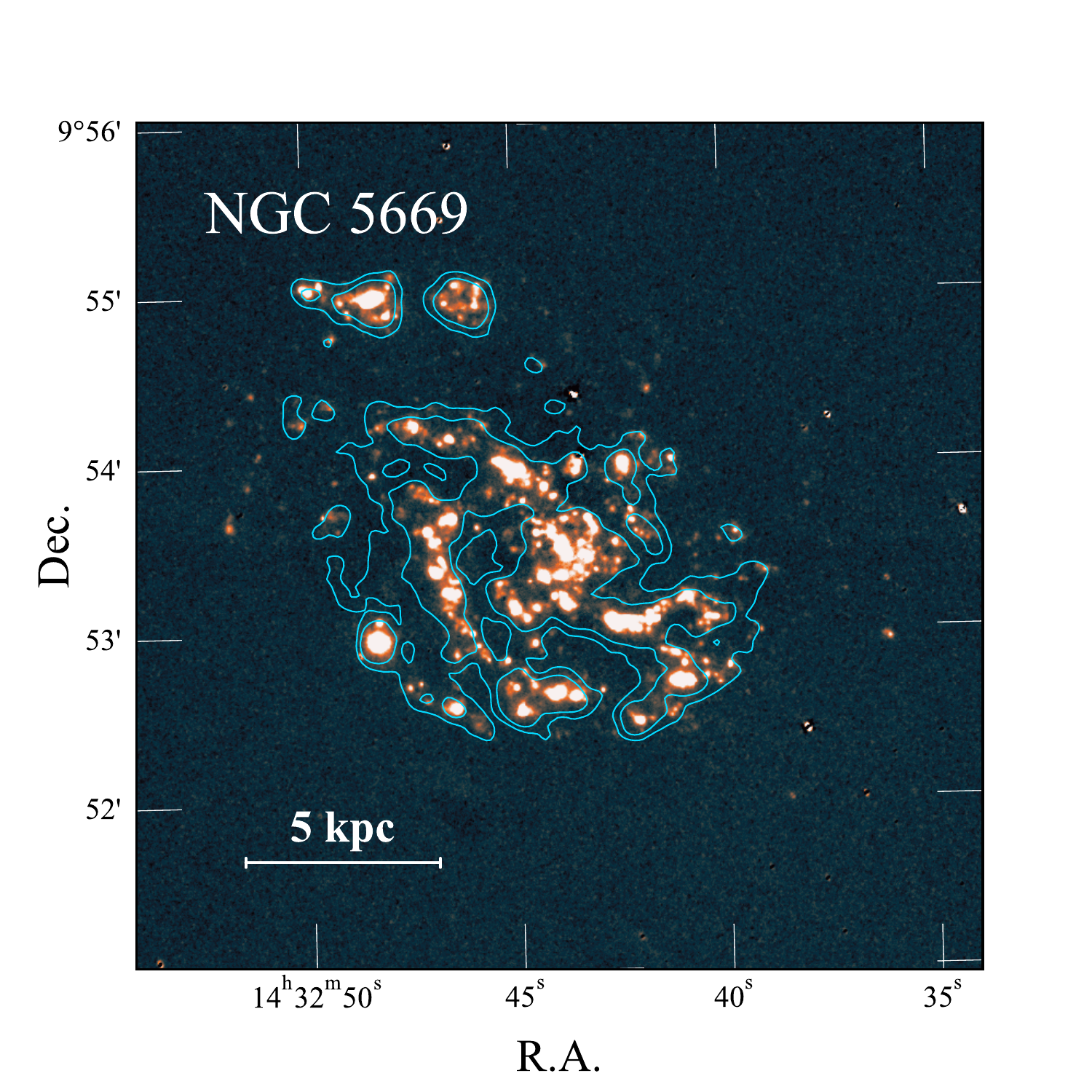}\hspace{-0.5cm}
\includegraphics[trim = 0cm 0cm 1cm 1cm, clip,scale=0.19]{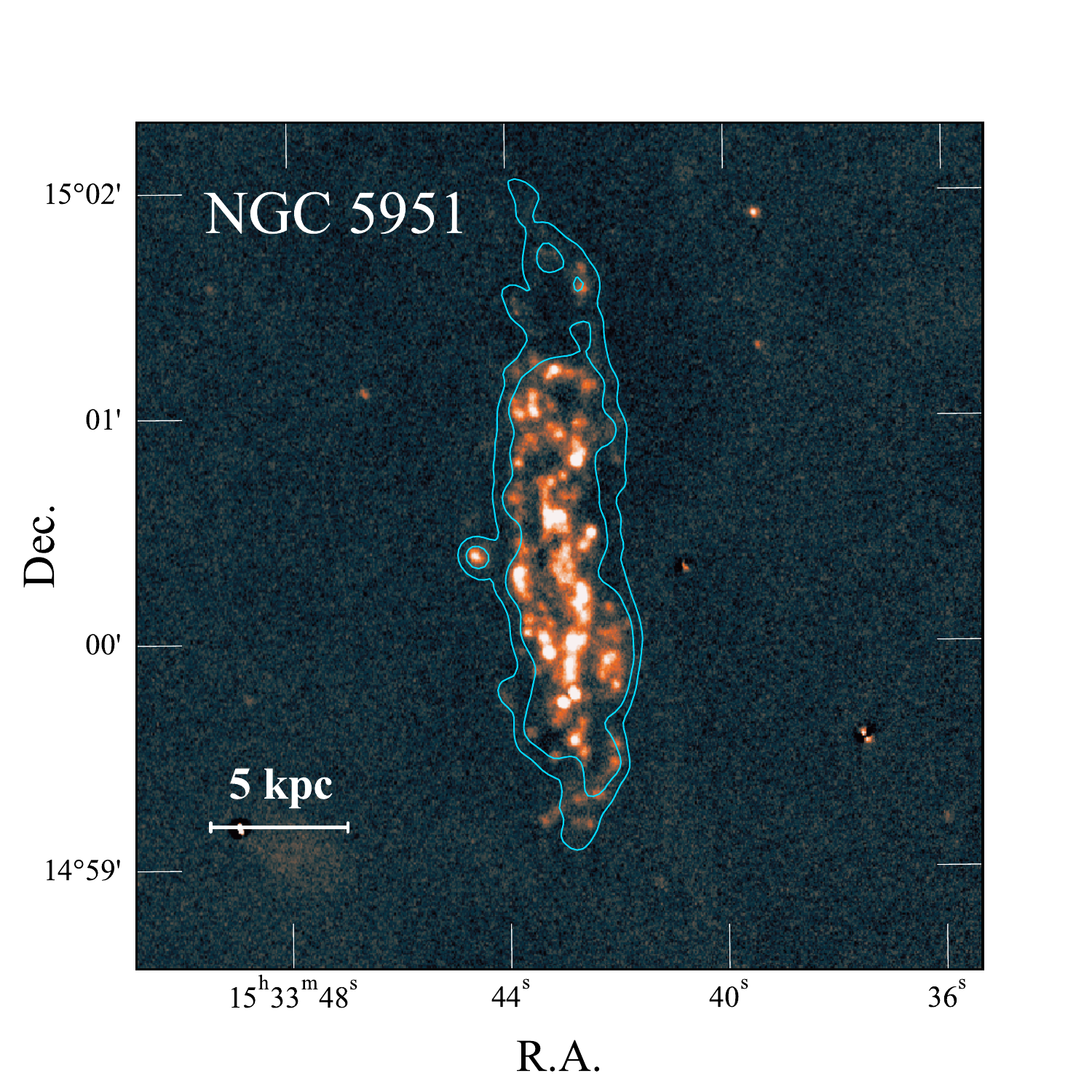}
\caption{Continuum-subtracted \ha\ imagery for our galaxies. FUV contours (blue) at 5$\sigma$ and 10$\sigma$ levels are overlaid on top. A physical scale of 5 kpc is shown in the bottom left corner of each map. North is up and East is towards left.}  \label{fig:hastampsuv}
\end{figure*}

The continuum-subtracted \hanii\ maps of our sample are shown in Figure \ref{fig:hastampsuv} with GALEX FUV contours overlaid on top. The sensitivity limits of the \hanii\ and FUV maps are noted in Table \ref{tab:gal}. In addition to the \hanii\ and FUV imaging, we use \wise\ and \mips\ images from the Wide-field Infrared Survey Explorer (WISE) All-Sky Data Release \citep{wright10} and Multiband Imaging Photometer for Spitzer \citep[MIPS;][]{mips}, respectively. 
All galaxies in our sample are also investigated in \cite{padave24}, hence, we refer the reader to that paper for details regarding the data mentioned above and additional sample properties.

\begin{table*}[!ht]
\caption{Properties of Galaxies in the Sample.}
\label{tab:gal}
\centering
\begin{tabular}{lcccccccccc}
\hline\hline
 \multirow{3}{*}{Galaxy} & \multirow{3}{*}{R.A.} & \multirow{3}{*}{Dec} & \multirow{3}{*}{cz} & \multirow{3}{*}{Distance} & \multirow{3}{*}{Scale}  & \multirow{3}{*}{$i$} & \multicolumn{2}{c}{Sensitivity Limit} & \multirow{3}{*}{SFR$_{\text{UV}}$} & \multirow{3}{*}{SFR$_{\text{H}\alpha}$} \\
        \cmidrule(lr){8-9}
        & & & & & & & FUV & \ha\ & & \\
 &  (deg) & (deg) & (km~s$^{-1}$) &  (Mpc) & (pc/\arcsec) & ($^\circ$) & ($\times10^{-19}$) & ($\times10^{-19}$) & ($M_\odot$~yr$^{-1}$) & ($M_\odot$~yr$^{-1}$)\\
 (1) & (2) & (3) & (4) & (5) & (6) & (7) & (8) & (9) & (10) & (11)\\
\hline
NGC 99  &  5.99747  &  15.77042  &  5313  &  79.4  &  378.17    &  20  & 9.05 & 7.21 & 2.45  & 1.48  \\  
NGC 3344  &  160.87951  &  24.92249  &  588  &  8.28  &  40.14    &  6  & 5.73 & 4.08 &  0.47  &  0.27  \\  
NGC 3351  &  160.99038  &  11.7038  &  778  &  9.29  &  45.04   &  44 & 4.51 & 2.59 &  0.79  & 0.44   \\  
NGC 3433  &  163.01611  &  10.1483  &  2724  &  44.59  &  216.18   &  22  & 4.92 & 6.03 & 1.20  & 0.46   \\  
NGC 3485  &  165.00991  &  14.84157  &  1433  &  29.43  &  142.68   &  23 & 4.45 & 4.40 &  0.83  &  0.59  \\  
NGC 3666  &  171.10861  &  11.34222  &  1060  &  17.1  &  82.90    &  75  & 15.40 & 11.30 &  0.32  & 0.25   \\ 
NGC 3810  &  175.24481  &  11.47112  &  993  &  15.3  &  74.18    &  47  & 18.50 & 4.36 &  1.25  &  0.48  \\  
NGC 4303  &  185.47894  &  4.47347  &  1567  &  18.7 &  90.66    &  22  & 6.05 & 6.96 &  5.25  & 6.05   \\  
NGC 4321  &  185.7289  &  15.82101  &  1576  &  13.93  &  67.53   &  24  & 2.87 & 5.74 &  2.49  &  1.46  \\  
NGC 5669  &  218.18311  &  9.8917  &  1369  &  14.5  &  70.30  &  48  & 16.50 & 2.78 &  0.29  & 0.17   \\  
NGC 5951  &  233.42933  &  15.00715  &  1778  &  27.1  &  131.38   &  77  & 8.45 & 6.23 &  0.23  & 0.11  \\  
\hline\hline 
\end{tabular}
\flushleft
  \begin{minipage}{\linewidth}
        \raggedright % Left-align the table notes
        \footnotesize % Smaller font size for table notes
        \begin{justify}
        \textbf{NOTES: }                  
        Column (1): Galaxies in our sample. Columns(2--3): J2000 celestial coordinates. Columns (4--6): recessional velocity, adopted distance, the physical scale at that distance in pc/\arcsec. Column (7): inclination estimated from the apparent axis ratio. Columns (8--9): 1$\sigma$ sensitivity limits of the GALEX FUV and VATT \ha\ data in erg~s$^{-1}$~cm$^{-2}$~\AA$^{-1}$ and erg~s$^{-1}$~cm$^{-2}$, respectively. Columns (10--11): Dust-corrected FUV and \ha\ SFRs.
        \end{justify}
    \end{minipage} 
\end{table*}

\subsection{Data Reduction \& Calibration of H$\alpha$ Images} \label{sec:hared}
%observation strategy and data reduction%
The VATT \hanii\ and \sdssr\ images were reduced as described in \cite{padave21} and \cite{padave24}. In brief, we use standard IRAF procedures to perform overscan correction, bias subtraction, and flat-field correction. 
Removal of cosmic rays was carried out using the \texttt{L.A.COSMIC} routine \citep{vand01}. 
Science exposure frames were aligned using \texttt{Astroalign} \citep{astroalign} before co-adding. The World Coordinate System of the final stacked images was refined with the \href{http://astrometry.net}{http://astormetry.net} software \citep{astrometry}. 
The background was estimated using Astropy Photutils \citep{larry_bradley_2023_7946442} and subsequently removed. Flux calibration of the \sdssr\ maps was carried out using the Aperture Photometry Tool \citep{apt} by matching the instrumental magnitudes to the SDSS DR12 magnitudes of field stars.
For the broad-band \sdssr\ images, we reach 1$\sigma$ depths of surface brightness of $\sim$25--27~mag~arcsec$^{-2}$. 

The \hanii\ images were processed further to remove the stellar continuum. The \sdssr\ image was appropriately scaled and then subtracted from the \ha\ image, to produce an emission-line only \hanii\ image. 
The scale factor, $F$, defined as the ratio of narrow-to-broad band counts, was estimated empirically using the field stars under the assumption that they do not emit \ha. 
We iterated the continuum subtraction process for a range of scale factors until the residual flux of non-saturated foreground stars was lowest, and the surface brightness of the continuum-dominated regions reached the the background level. 
This implicitly assumes the galaxy disk spectral energy distribution matches the foreground stars and does not vary with location. For the selected galaxies in our sample, the nominal scale factors were found to be between 0.031--0.047. 

The absolute flux measurements of the emission-only \hanii\ image were determined using the method outlined in \cite{kenn08}. We employed the photometric zero point (ZP) of the \sdssr-band image, the bandwidth (FWHM$_{\text{NB}}$) of the corresponding narrowband \ha\ filter, count rates (CR$_{\text{H}\alpha}$) in counts s$^{-1}$ from the continuum-subtracted \hanii\ map, and $F$ to compute the total emission-line flux, $f_{\text{tot}}$:
\begin{equation*}
    f_{\text{tot}} = U \cdot \text{FWHM}_{\text{NB}} \cdot \text{CR}_{\text{H}\alpha} \cdot T_{\lambda}^{-1}
\end{equation*}
Here, $U$ and $T_{\lambda}$ are calculated as follows:
\begin{align*}
    U&=\lambda^{-2}10^{-0.4[\text{ZP}+2.397-\kappa\sec(x)]}\\
    T_{\lambda}&=T_{NB}(\lambda) - T_{r}(\lambda)\frac{t_r}{t_{NB}}F
\end{align*}

We assumed a standard atmospheric extinction coefficient, $\kappa=0.08$~mag airmass$^{-1}$ for airmass $x$. The term $T_\lambda$ adjusts for the effective filter transmission at the redshifted \ha\ line, ensuring the recovery of the \ha\ flux from the \sdssr\ image lost during continuum subtraction. 
$T_{NB}(\lambda)$ and $T_{r}(\lambda)$ represent the normalized filter transmission, while $t_{NB}$ and $t_{r}$ denote the exposure times of the \ha\ and \sdssr\ maps, respectively. A Galactic (foreground) extinction correction is also applied based on \cite{schl11}, with A$_{H\alpha}=2.5\times E(\bv)$.  
We note that these calculations contain the contribution from the 
\nii$\lambda\lambda6548, 6584$ forbidden lines. All narrowband \ha\ observations of our sample include \nii\ emission and the process of subtracting the \nii\ flux and its implications are discussed in \S\ref{sec:sfr}.

\section{FUV- \& H$\alpha$-bright Stellar Complexes}\label{sec:reg}
\subsection{Source Detection}
We created a catalog of both UV- and \ha-bright objects in our galaxies using \texttt{Source Extractor} \citep{bert96} and its Python library, \texttt{sep} \citep{sep}. UV-bright stellar complexes or star-forming clumps are detected using the GALEX maps and are referred to as FUV-selected regions. \hii\ regions, \ha\ knots or \ha-bright objects detected using the VATT \hanii\ maps are termed as \ha-selected regions.

The FUV-selected regions were detected by extracting sources 3$\sigma$ above the background level of the FUV maps. To optimize source extraction in various environments, such as the densely populated central regions of galaxies and the sparser galactic outskirts, we execute separate runs of \texttt{sep} on cutouts of different regions within each galaxy. Adjustments to the background box and filter sizes, as well as the deblending threshold and contrast parameter, were made for each run to ensure efficient object detection and minimize overlap between detected objects. After a preliminary catalog of objects is created for a galaxy, the catalog is cleaned for background and foreground contaminants and overlapping apertures. Apertures at the positions of foreground stars seen in the optical images were removed. Additionally, manual adjustments were made to redefine certain apertures to encompass detected regions effectively.

A similar process is employed for extracting \ha-selected regions from the \hanii\ maps. Prior to source extraction, the \hanii\ maps, with a resolution of $\sim$1\farcs5 FWHM, are convolved to match the resolution of the GALEX FUV map (4\farcs2 FWHM). To achieve this, the \hanii\ maps, which have a pixel scale of 0\farcs375/pixel, were first resampled to 1\farcs5/pixel to match the FUV maps and then smoothed. This process ensures consistent source detection and spatial matching, facilitating a direct comparison between the FUV-selected and \ha-selected regions.

For the final catalog, we also ensure that only regions reliably detected in both the \ha\ and FUV datasets are considered. For this, a region is included in the final catalog only if it is detected above the sensitivity limits of both maps. The final catalog comprises 1335 regions detected in the FUV and 1474 regions detected in \ha\ across our sample of 11 DIISC galaxies. In Figure \ref{fig:histogram}, we show histograms depicting the distributions of various properties of the  FUV- and \ha-selected regions, along with their radial distribution. Properties of the regions detected in each galaxy are described in the following sections and summarised in Table \ref{tab:reg}.

\begin{table*}[!ht]
\caption{Properties of the Detected FUV- and \ha-selected Sources}
\label{tab:reg}
\centering
\renewcommand{\arraystretch}{0.9}
    \begin{tabular}{lcccccccccc}
        \hline\hline
        \multirow{2}{*}{Galaxy} &
        \multirow{2}{*}{R$_{25}$} &
        \multirow{2}{*}{Adopted [N\,\small{II}]/H$\alpha$} &
        \multicolumn{4}{c}{FUV-selected Regions} &
        \multicolumn{4}{c}{H$\alpha$-selected Regions} \\
        \cmidrule(lr){4-7} \cmidrule(lr){8-11}
        & &  value or Gradient & N & Sizes & Areas & $R_{\text{max}}$ & N & Sizes (kpc) & Areas & $R_{\text{max}}$ \\
        & (kpc) & & & (kpc) & (kpc$^2$) & ($R_{25}$) & & (kpc) & (kpc$^2$) & ($R_{25}$) \\
        (1) & (2) & (3) & (4) & (5) & (6) & (7) & (8) & (9) & (10) & (11)\\
\hline
\multirow{2}{*}{NGC 99} & \multirow{2}{*}{11.46} & \multirow{2}{*}{$-0.03R_{25}-0.56$} & \multirow{2}{*}{27} & 1.04 -- & 2.50 -- & \multirow{2}{*}{1.56} & \multirow{2}{*}{21} & 0.76 -- & 1.16 -- & \multirow{2}{*}{1.50} \\
    & & & & \raggedleft 3.63 & \raggedleft 15.24 & & & \raggedleft 3.33 & \raggedleft 16.79 & \\
    \hline
    \multirow{2}{*}{NGC 3344} & \multirow{2}{*}{7.93} & \multirow{2}{*}{$-0.91R_{25}+0.77$} & \multirow{2}{*}{315} & 0.09 -- & 0.02 -- & \multirow{2}{*}{1.24} & \multirow{2}{*}{312} & 0.08 -- & 0.02 -- & \multirow{2}{*}{1.24}\\
    & & & & \raggedleft 0.66 & \raggedleft 0.80 & & & \raggedleft 0.49 & \raggedleft 0.38 &\\
    \hline
    \multirow{2}{*}{NGC 3351} & \multirow{2}{*}{8.15} & \multirow{2}{*}{$-0.14R_{25}-0.47$} & \multirow{2}{*}{230} & 0.09 -- & 0.01 -- & \multirow{2}{*}{1.46} & \multirow{2}{*}{261} & 0.07 -- & 0.01 -- & \multirow{2}{*}{1.42}\\
    & & & & \raggedleft 0.55 & \raggedleft 0.47 & & & \raggedleft 0.37 & \raggedleft 0.28 &\\
    \hline
    \multirow{2}{*}{NGC 3433} & \multirow{2}{*}{17.66 }& 0.67 ($R\leq R_{25}$) & \multirow{2}{*}{102} & 0.53 -- & 0.63 -- & \multirow{2}{*}{1.67} & \multirow{2}{*}{105} & 0.42 -- & 0.39 -- & \multirow{2}{*}{1.52}\\
    & &  0 ($R>R_{25}$) & & \raggedleft 2.36 & \raggedleft 8.25 & & & \raggedleft 1.64 & \raggedleft 4.38 &\\
    \hline
    \multirow{2}{*}{NGC 3485} & \multirow{2}{*}{9.87} & 0.58 ($R\leq R_{25}$) & \multirow{2}{*}{62} & 0.37 -- & 0.29 -- & \multirow{2}{*}{1.84} & \multirow{2}{*}{78} & 0.25 -- & 0.16 -- & \multirow{2}{*}{1.85}\\
    & & 0 ($R>R_{25}$) & & \raggedleft 1.26 & \raggedleft 3.18 & & & \raggedleft 0.86 & \raggedleft 1.36 &\\
    \hline
    \multirow{2}{*}{NGC 3666} & \multirow{2}{*}{4.90} & 0.56 ($R\leq R_{25}$) & \multirow{2}{*}{29} & 0.23 -- & 0.14 -- & \multirow{2}{*}{1.86} & \multirow{2}{*}{33} & 0.20 -- & 0.10 -- & \multirow{2}{*}{2.18}\\
    & & 0.13 ($R> R_{25}$) & & \raggedleft 0.84 & \raggedleft 1.22 & & & \raggedleft 0.59 & \raggedleft 0.73 &\\
    \hline
   \multirow{2}{*}{NGC 3810} & \multirow{2}{*}{7.54 }& 0.73 ($R\leq R_{25}$)& \multirow{2}{*}{61} & 0.22 -- & 0.10 -- & \multirow{2}{*}{1.14} & \multirow{2}{*}{94} & 0.14 -- & 0.04 -- & \multirow{2}{*}{1.42}\\
    & & 0.29 ($R> R_{25}$)& & \raggedleft 1.05 & \raggedleft 1.69 & & & \raggedleft 0.64 & \raggedleft 0.67 &\\
    \hline
    \multirow{2}{*}{NGC 4303} & \multirow{2}{*}{17.37} & \multirow{2}{*}{$-0.74R_{25}-0.53$} & \multirow{2}{*}{223} & 0.17 -- & 0.08 -- & \multirow{2}{*}{1.10} & \multirow{2}{*}{207} & 0.15 -- & 0.06 -- & \multirow{2}{*}{1.10}\\
    & & & & \raggedleft 1.02 & \raggedleft 1.69 & & & \raggedleft 0.83 & \raggedleft 1.04 &\\
    \hline
    \multirow{2}{*}{NGC 4321} & \multirow{2}{*}{12.85} & \multirow{2}{*}{$-0.16R_{25}-0.61$} & \multirow{2}{*}{219} & 0.14 -- & 0.06 -- & \multirow{2}{*}{1.03} & \multirow{2}{*}{270} & 0.11 -- & 0.03 -- & \multirow{2}{*}{1.16 }\\
    & & & & \raggedleft 1.07 & \raggedleft 1.29 & & & \raggedleft 0.53 & \raggedleft 0.68 &\\
    \hline
    \multirow{2}{*}{NGC 5669} & \multirow{2}{*}{6.52} & 0.36 ($R\leq R_{25}$) & \multirow{2}{*}{53} & 0.21 -- & 0.11 -- & \multirow{2}{*}{1.45} & \multirow{2}{*}{72} & 0.13 -- & 0.05 -- & \multirow{2}{*}{1.42}\\
    & & 0 ($R>R_{25}$) & & \raggedleft 0.60 & \raggedleft 0.73 & & & \raggedleft 0.67 & \raggedleft 1.05 &\\
    \hline
    \multirow{2}{*}{NGC 5951} & \multirow{2}{*}{6.88} & 0.49 ($R\leq R_{25}$) & \multirow{2}{*}{15} & 0.37 -- & 0.24 -- & \multirow{2}{*}{1.90} & \multirow{2}{*}{22} & 0.32 -- & 0.26 -- & \multirow{2}{*}{1.88} \\
    & & 0 ($R > R_{25}$) & & \raggedleft 1.49 & \raggedleft 4.81 & & & \raggedleft 1.10 & \raggedleft 1.84 & \\
        \hline\hline
\end{tabular}
\flushleft
  \begin{minipage}{\linewidth}
        \raggedright % Left-align the table notes
        \footnotesize % Smaller font size for table notes
        \begin{justify}
        \textbf{NOTES: }                  
        Column (1): Galaxies Name. Column(2): Radius at which the galaxy's surface brightness drops to 25 mag/\arcsec$^2$ in the B-band. The estimates are taken from \cite{padave24}. Column (3): Adopted \nii/\ha\ values or \nii/\ha\ radial gradients. See details in \S\ref{sec:sfr}. Columns (4--7):  Number of FUV-selected regions, their range of physical sizes in kpc and surface areas in kpc$^2$, and distance of the furthest detected object normalized by $R_{25}$ for the galaxy. Column (8--11): Same as Columns (4--7) but for \ha-selected regions.
        \end{justify}
    \end{minipage} 
\end{table*}

\subsection{Source Sizes \& Morphologies}\label{sec:morph}

\begin{figure*}[!t]
\centering
\includegraphics[trim =0cm 0cm 5cm 0cm, clip,scale=0.45]{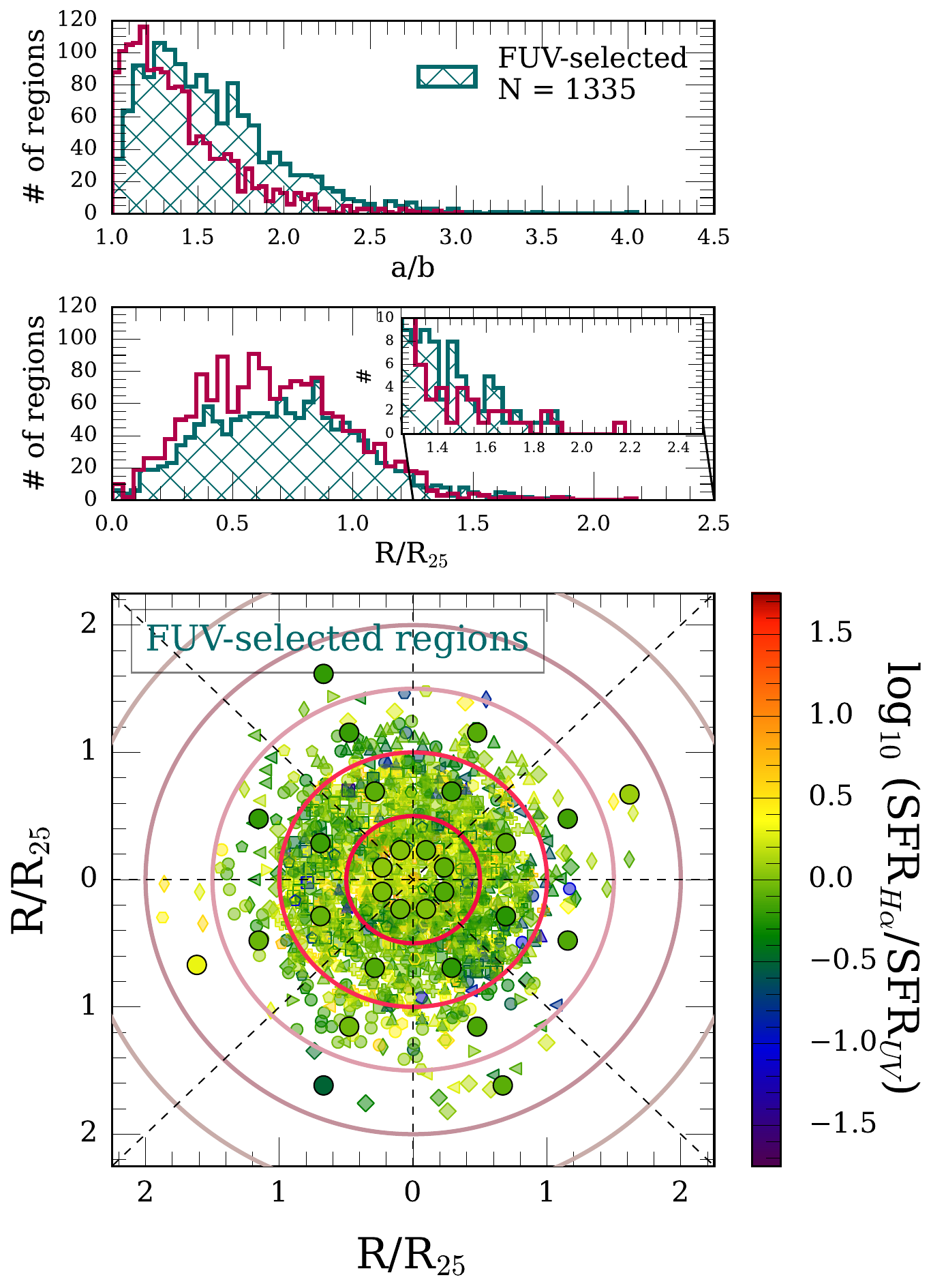}
\includegraphics[trim = 2.6cm 0cm 0cm 0cm, clip,scale=0.45]{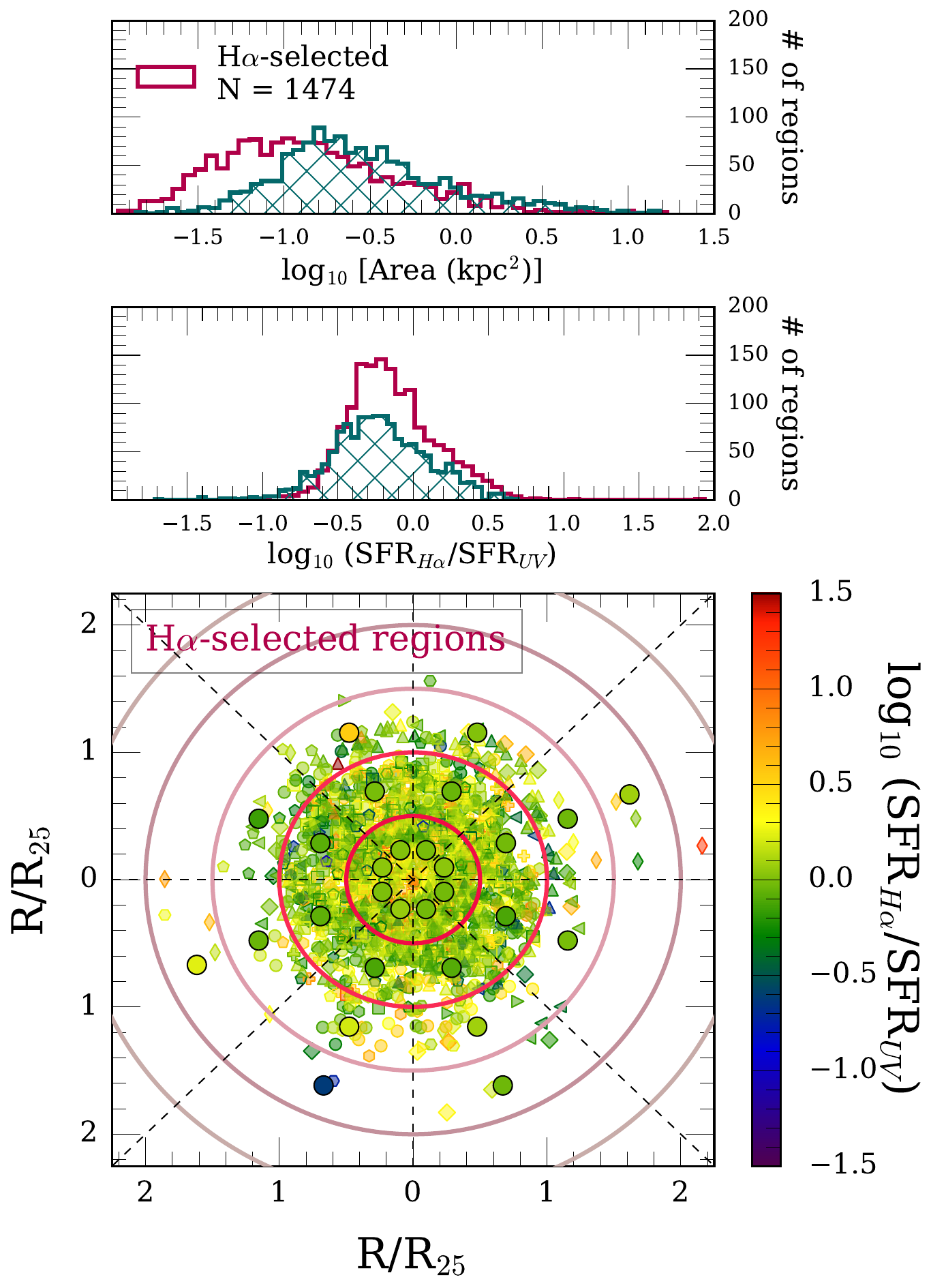}
\caption{Histograms showing the distribution of (top) axis ratios ($a/b$) for elliptical apertures, their surface areas, (middle) deprojected $R/R_{25}$, and the \sfrha/\sfruv\ ratio of the FUV- and \ha-selected regions. The inset in the middle left panel shows a zoom-in of the histogram for $1.25R_{25}\leq R\leq 2.5R_{25}$. The bottom panels show the locations of the FUV-selected (left) and \ha-selected (right) regions on the $R/R_{25}$ plane for a representative galaxy. Each region is color-coded based on its \sfrha/\sfruv\ ratio with unique symbols used in Figure \ref{fig:glob_sfr} to identify the sample galaxy. The symbols are not displayed on the plot for brevity. The overlaid larger circles show the average \sfrha/\sfruv\ in a given azimuthal and radial bin marked by the dashed lines and solid circles, only shown for bins with more than one region. The concentric circles show $R=0.5, 1, 1.5, 2, 2.5\times R_{25}$. } \label{fig:histogram}
\end{figure*}

The physical sizes of the detected regions vary between 60~pc to 3.6~kpc. These range from compact, isolated stellar associations or \hii-regions to UV-bright star-forming clumps or conglomerations of \hii\ regions. Approximately 70\% of the \ha-selected regions have sizes ranging from  60--320~pc,  indicating a comparatively more compact size than the FUV-selected regions, where around 60\% of the regions range from 150--500~pc. 
Consequently, the morphology of the FUV-selected regions tends to be more elongated than the \ha-selected regions, as observed in the top left panel of Figure \ref{fig:histogram}. The axis ratios, $a/b$, of the elliptical apertures reveal that $\sim$50\% of the FUV-selected regions have $a/b$ $>$ 1.5, while $\sim$70\% of the \ha-selected regions are compact. The areas of the \ha-selected regions are also smaller than those of the FUV-selected regions, as depicted in the top right panel of Figure \ref{fig:histogram}. Unsurprisingly, the larger regions with areas $\gtrsim$1~kpc$^2$ in the catalog are predominantly from the more distant galaxies in our sample: NGC~99, NGC~3433, and NGC~3485, as a direct consequence of degradation of resolution. 

\subsection{Radial Distribution of Sources}\label{sec:rad}
We observe an almost identical radial distribution of the FUV- and \ha-selected objects across our sample, as illustrated in the middle left panel of Figure \ref{fig:histogram}. The deprojected galactocentric distances of the regions, $R$, are normalized by $R_{25}$ of the galaxy, as taken from \cite{padave24}. The 2D distribution of detected regions is shown in the bottom panel. We note that this panel is mainly to illustrate the distribution of sources beyond $R_{25}$. Approximately 80\% of the sources lie within $R_{25}$ and we detect 248/1335 FUV- and 222/1474 \ha-selected regions beyond the optical disk. About 22\% of these outer objects were detected in the extended ultraviolet (XUV) disk of NGC~3344. 
The farthest UV clump in our catalog at 1.9$R_{25}$ in NGC~5951 was also detected in \ha. The farthest \ha-selected region was detected in NGC~3666 at 2.18$R_{25}$ but is not detected in FUV. We note that these galaxies are highly inclined, and deprojection issues are unavoidable. 
However, as we already take into account the detection thresholds of both the FUV and \ha\ maps, regions that are undetected by GALEX could be very young, such that these complexes may still be embedded within their birth cloud and thus highly obscured.

\subsection{H$\alpha$-to-FUV Ratios of Sources}\label{sec:sfr}
The \ha-to-FUV SFR ratios, or \sfrha/\sfruv\ are estimated for all FUV- and \ha-selected regions. Before estimating the SFRs, we attempt to retrieve the true \ha\ flux from the \hanii\ maps. Corrections for the contamination from \nii$\lambda\lambda6548, 6584$ forbidden lines are implemented using \nii/\ha\ radial gradients or values sourced from the literature. These are noted in Table \ref{tab:reg} and described below. 

Olvera et al. (2024) investigated the optical spectra of \hii\ regions in NGC 99, estimating the metallicity gradient in the galaxy using several emission lines-selected indicators, including \nii/\ha. They found that the \nii\ flux is $\sim$25\% of \ha\ in the center, decreasing to $\sim$6\% in the outer zones of the galaxy. 
Similarly, in NGC~3344, the \nii/\ha\ ratios range from 0.6 in the inner regions to 0.2 in the outer regions \citep{moum19}. We make use of the \nii/\ha\ gradient from \cite{moum19} to recover the \ha\ flux in NGC~3344 in this study instead of the integrated \nii/\ha\ value of 0.52 used in \cite{padave21}. 
For NGC~3351, NGC~4303, and NGC~4321, gas-phase metallicity gradients from \cite{pilg23} were used along with calibrations provided in \cite{curti20}, to derive the \nii/\ha\ gradient in these galaxies. \nii/\ha\ estimates for other galaxies were obtained from SDSS Data Release 18 \citep{sdssdr18}, and for NGC 3485, the ratio was sourced from \cite{ho97}. Given that \nii\ fluxes typically dominate the metal-rich inner disks of galaxies and decrease with R, we adopt different values for regions within $R_{25}$ compared to those beyond and reduce systematic uncertainties. 
For most galaxies, the \nii/\ha\ estimate is derived from the nuclear region spectra in SDSS and is used to remove the contribution from regions with $R\leq R_{25}$. In cases where estimates for regions with $R > R_{25}$ were available, those values were used for \nii/\ha; otherwise, no contribution from \nii\ was assumed to avoid any discrepancies due to varying metallicity gradients among galaxies \citep[][Olvera et al. 2024]{pilg23}. 

We also account for internal dust extinction of the FUV and \ha\ fluxes, which have only been corrected for foreground extinction so far. 
For extinction correction, a combination of FUV and \ha\ with infrared fluxes has been widely used and provides both the unobscured and dust-obscured components of the star formation \citep{calz07, kenn07}. 
We use the following calibrations computed in \cite{kenn98}, \cite{calz07} and \cite{lero08} to calculate the inherent \sfruv\ and \sfrha\ of the detected regions:
 \begin{align*}
      SFR_{\rm UV} &= 0.68\times10^{-28}~{\rm L}_{\nu, \rm{FUV}} + 2.14\times10^{-42}~{\rm L}_{\rm{24}\mu{\rm m}}\\
 SFR_{\rm{H}\alpha} &= 5.3\times10^{-42}(\rm{L}_{\rm{H}\alpha}+3.1\times10^{-2}~\rm{L}_{\rm{24}\mu{\rm m}})
\end{align*}
where \sfruv\ and \sfrha\ are in M~$_\odot$~yr$^{-1}$, and the FUV, \ha, and \mips\ luminosity, L$_{\nu, \rm{FUV}}$, L$_{H\alpha}$, and L$_{24\mu m}$ are  in erg~s$^{-1}$~Hz$^{-1}$, erg~s$^{-1}$, and erg~s$^{-1}$, respectively. The \sfruv\ and \sfrha\ prescriptions assume a \cite{salp55} and \cite{kroupa01} initial mass function and a factor of 1.59 is used for the conversion. While L$_{\nu, \rm{FUV}}$ and L$_{H\alpha}$ take no account of internal extinction, L$_{24\mu m}$ is assumed to account for the dust-processed light.
Where \mips\ data is not available, we use \wise\ fluxes. 

We only apply a first-order extinction correction from the IR maps to a detected region when estimating SFRs instead of estimating \wise\ or \mips\ luminosities for individual regions. Firstly, the \wise\ observations have a resolution of 12\arcsec, which is significantly worse than our FUV and \ha\ maps. 
As we are mainly interested in investigating UV- and \ha-bright star-forming clumps or \hii\ regions, convolving these maps to 12\arcsec\ would result in missing out on low-surface brightness regions in the outskirts and other structural details. 
Secondly, the \mips\ observations, especially beyond $R_{25}$, have low signal-to-noise ratios. Hence, any attempt to perform photometry on the IR maps would result in additional uncertainties in our SFRs. 
Instead, we make use of the \mips\ and \wise\ surface brightness profiles from \cite{padave24} to estimate the mean IR value at a given R. This mean value is then incorporated in the second term of the SFR equations for a region at a given R.

In the middle right panel of Figure \ref{fig:histogram}, we present the histogram of \sfrha/\sfruv\ for the regions. 
We find that the FUV-selected regions exhibit an almost symmetric and broader distribution, with most ratios falling between 0.34--0.9. The \ha-selected regions generally have \sfrha/\sfruv\ ratios $\gtrsim$0.45, with more regions showing \sfrha$>$\sfruv. Overall, FUV-selected regions tend to display lower \sfrha/\sfruv\ compared to \ha-selected regions. In the 2D distribution shown in the bottom panels of Figure \ref{fig:histogram}, the regions are colored according to $\log_{10}$~\sfrha/\sfruv. The average ratios per azimuthal and radial bin are over-plotted as large circles for convenience in bins where the number of regions is more than 1. Not much variation in the \ha-to-FUV ratios is observed in the bins. Additionally, we find a considerable scatter of 0.36 dex and 0.33 dex per annular bin in the ratios in the FUV- and \ha-selected \ha-to-FUV ratios, respectively. We explore radial trends of the SFRs and the ratios in the following section.

\section{Results} \label{sec:res}

\subsection{Radial Trends in SFRs}\label{sec:radpro}

\begin{figure*}[!ht]
  \centering
\includegraphics[trim = 3.5cm 0cm 5.cm 3cm, clip,scale=0.365]{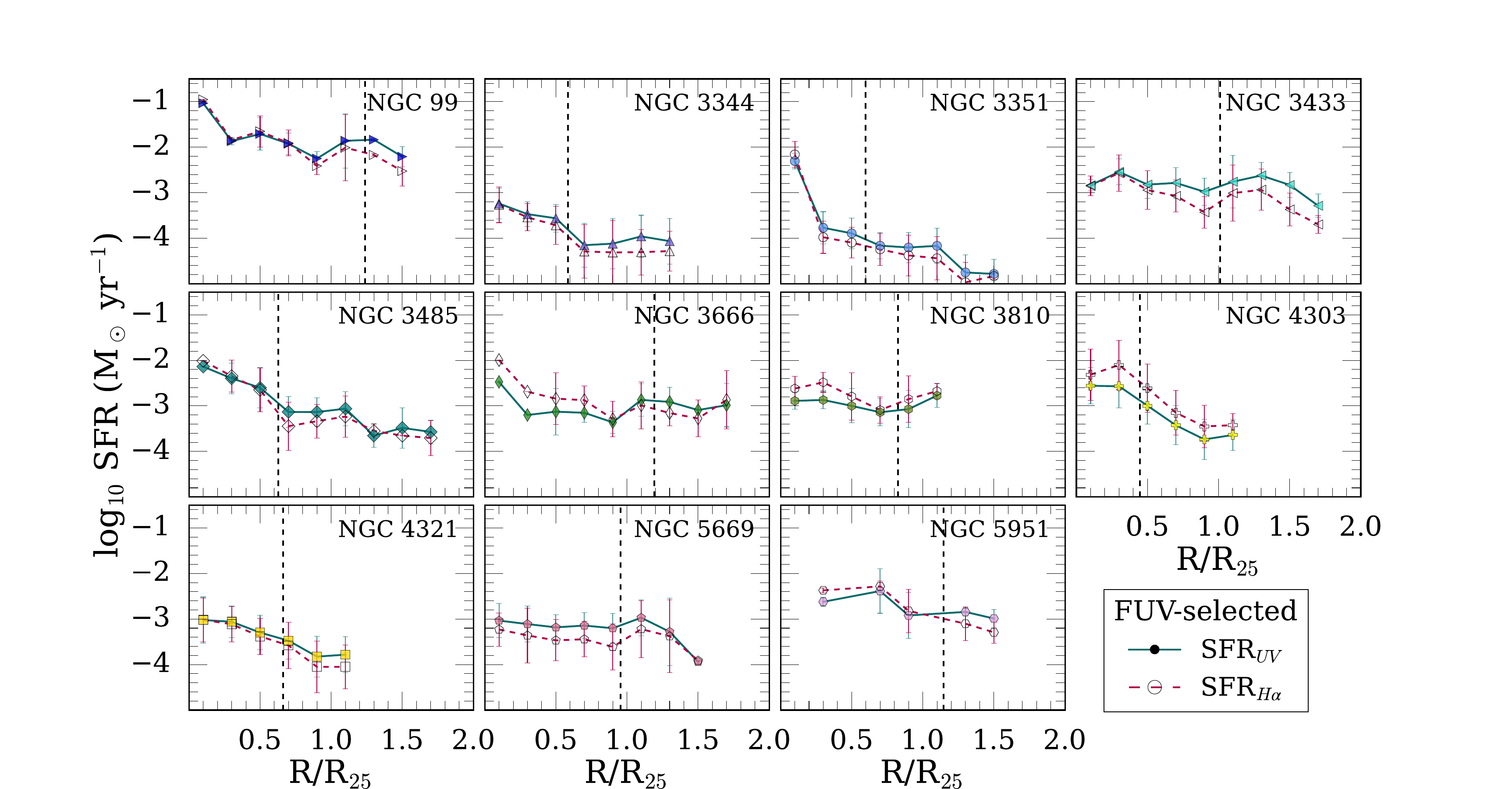}
\caption{Radial variation in \sfruv\ (solid lines) and \sfrha\ (dashed lines) for the FUV-selected regions in each galaxy in our sample. The filled and unfilled symbols denote the average \sfruv\ and \sfrha\, respectively, within a radial bin. The 1$\sigma$ scatter in the SFRs at each radius is denoted by the vertical bars. 
The vertical dashed line shows the radii containing 90\% of the \sfruv\ in the galaxy, taken from \cite{padave24}.} \label{fig:fradpro}
\end{figure*}

\begin{figure*}[!ht]
  \centering
\includegraphics[trim = 3.5cm 0cm 5.cm 3cm, clip,scale=0.365]{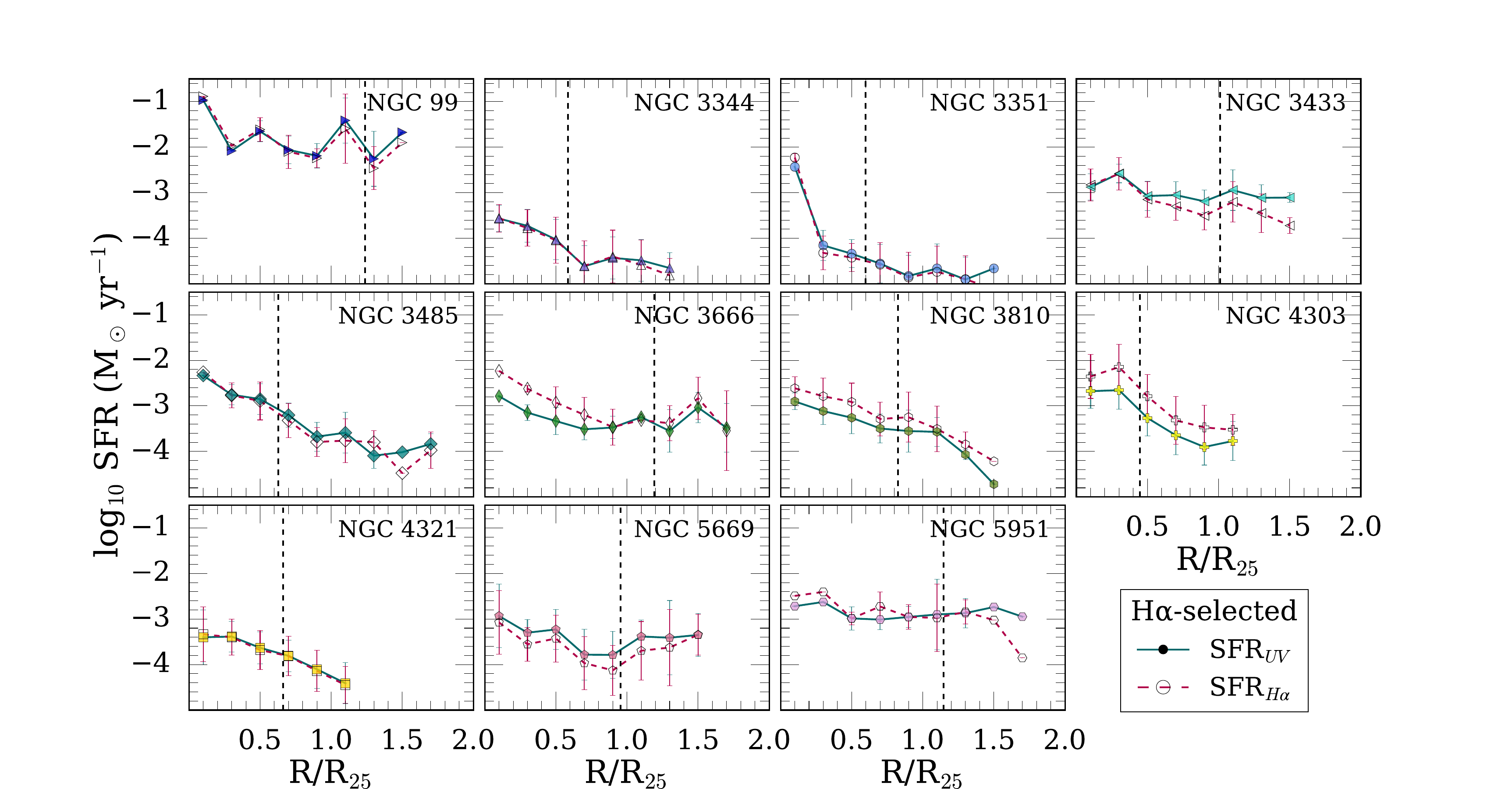}
\caption{Same as Figure \ref{fig:fradpro} but for the \ha-selected regions in each galaxy in our sample.} \label{fig:hradpro}
\end{figure*}

In Figures \ref{fig:fradpro} and \ref{fig:hradpro}, we show the variation in \sfruv\ and \sfrha\ as a function of $R/R_{25}$ for the FUV-selected and \ha-selected regions, respectively. In each panel, average \sfruv\ and \sfrha\ within a radial bin are estimated from the detected regions and denoted by filled and unfilled symbols, respectively, for a galaxy. The scatter in the SFRs in each bin is represented by the vertical bars. For reference, we show the radii containing 90\% of the SFR estimated from \sfruv\ distribution in \cite{padave24} as the vertical dashed line.

In Figure \ref{fig:fradpro}, we observe a general decline in \sfruv\ and \sfrha\ with $R/R_{25}$ in most galaxies for the FUV-selected regions. \sfrha\ is typically lower compared to \sfruv, especially at large galactocentric distances. The SFRs are seen to closely follow each other within $0.5 R_{25}$ and then begin to deviate beyond $0.5 R_{25}$. 
Notably, NGC 3810, and NGC 4303 do not follow these trends, showing \sfrha\ surpassing \sfruv. We bring to the reader's attention that dust correction for our SFRs was done using annular-averaged \mips\ or \wise\ value. 
Additionally, the poor resolution and depths of the \wise\ maps are also causing the IR fluxes to be spread out over areas much larger than the regions we are probing. This may be affecting these galaxies and some of the regions, where \sfruv\ is observed to be lower than \sfrha. 

We also note no evidence of truncation in the \ha\ emission.
Sharp turnovers in \ha\ surface density profiles around $R_{25}$ have been observed in many galaxies and attributed to gravitational, gas-phase, or cloud fragmentation instabilities and even truncation of the upper-mass end of the IMF \citep{mart01, scha04, krum08, pflamm08}. The absence of truncation in our sample is likely because we are not considering surface densities but rather the average SFR of all detected regions in a given bin. 

From the observed offsets in the FUV-selected regions across the sample, we find that the \ha-to-FUV SFR ratios generally decline with $R/R_{25}$. We note $\log_{10}$ \sfrha/\sfruv\ to decrease by 0.3$\pm0.03$~dex/$R_{25}$ on average with Pearson-$r$ value of -0.27. This weak dependence on $R/R_{25}$ highlights the fact that the radial trends for individual galaxies also show variations and exhibit considerable scatter. As the number of regions per galaxy shows a large range with the lowest value of only 15 detected FUV-selected regions in NGC~5951, we refrain from fitting radial trends for individual galaxies and look at the sample as a whole. 
The decline in \ha-to-FUV ratios, albeit weak, suggests that the \sfrha\ of individual regions decreases rapidly compared to \sfruv\ in the galactic outskirts. If this scenario is entirely true, then we would expect similar radial trends and offsets for the \ha-selected regions.

Interestingly, the \ha-selected profiles in Figure \ref{fig:hradpro} present a different picture. Here, both \sfruv\ and \sfrha\ closely follow each other in the inner parts without significant offsets as those observed in the FUV-selected profiles. The SFRs begin to deviate around $R_{25}$ but the decline in \sfrha\ is still less pronounced compared to the FUV-selected trends.  
This indicates that where \ha\ knots are detected with UV counterparts, the \ha\ fluxes are generally comparable to the UV fluxes. Similar observations were noted by \cite{godd10} in XUV disk galaxies, where \hii\ regions in the outskirts showed comparable \ha-to-FUV SFR ratios to those in the inner disks. Consequently, the \ha-to-FUV SFR ratios for our \ha-selected regions are noted to decrease by only 0.15$\pm$0.02 dex/$R_{25}$ for the entire sample and show no dependence on $R/R_{25}$ (Pearson-$r$ of -0.15).  This difference in trends noted between the FUV- and \ha-selected regions suggests that the decline in the \ha\ emission compared to the FUV does not just result from differences in declining fluxes. 

The fact that the extent of the discrepancies in the SFRs varies in the two cases also underscores fundamental differences in the star-forming regions probed by the two indicators. 
We perform the Kolmogorov-Smirnov (KS) test on the distribution of FUV-selected \sfruv\ and \ha-selected \sfrha, under the null hypothesis that the two samples are drawn from the same parent distribution. 
We obtain a $p$-value~$\lesssim0.01$ and hence, find that the two distributions are inherently different. This is expected as UV and \ha\ complexes represent different physical star-forming phenomena. While the \ha\ emission arises from \hii\ regions fueled by OB stars, the FUV complexes can encompass multiple \hii-regions, runaway OB stars, and longer-lived late-type field B- and A-stars. These lead to morphological differences between the two, where the FUV-selected regions are more elongated and dispersed than the compact \ha-complexes, also impacting \ha-to-FUV SFR ratios. Hence, differences in the FUV and \ha\ apertures as well as morphologies of the regions are playing a significant role. We explore this further in \S\ref{sec:overlap}.

\subsection{Spatial Distribution \& Overlaps of FUV and H$\alpha$ Emission}\label{sec:overlap}

The observations in the previous section allude to aperture mismatches between the FUV-selected detections and the morphology of \hii\ regions. The larger FUV-selected apertures compared to the concentrated morphologies of \ha-selected regions, as discussed in \S\ref{sec:morph}, cause \sfrha\ to be generally lower for FUV-selected detections. In larger apertures, \ha\ emission originates only from the portion of the aperture containing \hii\ regions. The aperture may include diffuse \ha\ emission, but it is likely picking up noise in areas where UV emission is dominant. Additionally, UV clumps may also not have any associated \hii\ regions \citep{gil05, thilk05, zaritsky07}. 

\begin{figure*}[!t]
  \centering
\includegraphics[trim = 0.cm 0cm 0cm 0.cm, clip,scale=0.4]{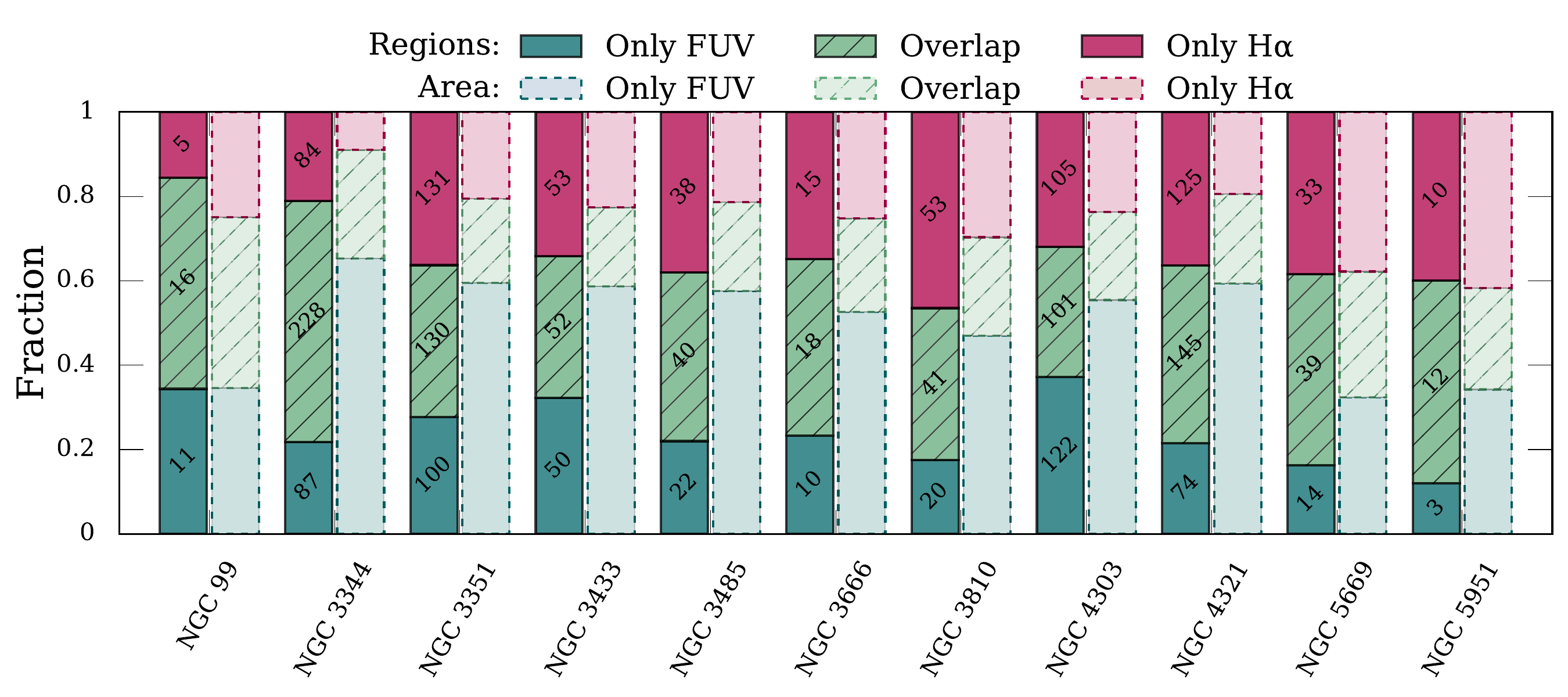}
\caption{Stacked bar plots showing the fraction of detected regions (left) and their corresponding areas (right) for each galaxy. Each bar is subdivided into three categories: regions detected only in FUV (bottom), overlapping regions, i.e., regions detected in both FUV and \ha\ (middle), and regions detected only in \ha\ (top). The number of regions associated with each category for each galaxy is noted on the bar on the left.} \label{fig:bar}
\end{figure*}

In Figure \ref{fig:bar}, we present two stacked bar plots for each galaxy: the first bar shows the number of regions, and the second shows the area associated with the regions. The bar plots are further subdivided into three categories: only FUV, both FUV and \ha, and only \ha\ regions. The number of detected regions and areas for each category are normalized by their corresponding totals for each galaxy. A region is considered to be detected in both FUV and \ha\ if the FUV- and \ha-selected apertures overlap by at least 5 pixels, and are denoted as overlap in the plot. Otherwise, the regions are considered only FUV or only \ha\ detections.

The areal contrast between the FUV-selected and \ha-selected detections is clearly illustrated in Figure \ref{fig:bar}. FUV regions occupy 55--85\% of the total detected area in our galaxies. Further, a significant number of regions, roughly 30\%, are detected in both FUV and \ha, covering $\sim$20\% of the total detected area. The \ha-selected regions cover a lower detected area of 25--65\% than FUV. However, most galaxies have more \ha-only regions compared to FUV-only. This is not surprising, as \ha\ distributions are clumpier as they are directly linked to the presence of \hii\ regions and easily detected. However, it is plausible that the continuum UV emission associated with the \hii\ region is diffuse and widespread, and doesn't necessarily show clumping to be detected. Accordingly, the FUV-only detections imply that these regions do not contain massive stars capable of producing \hii\ regions.

\begin{figure*}[!ht]
  \centering
\includegraphics[trim =3cm 0cm 0cm 2.8cm, clip,scale=0.355]{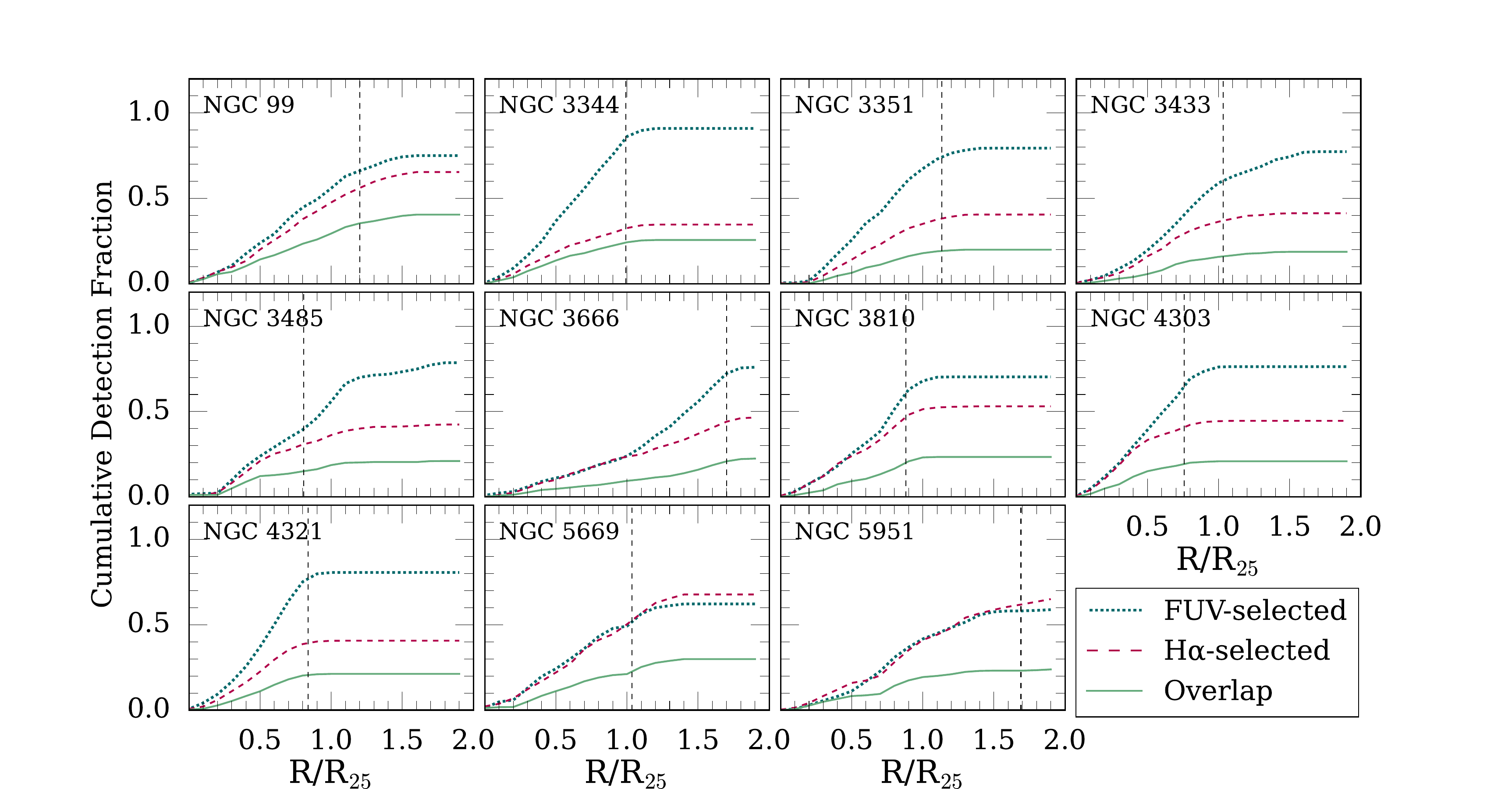}
\caption{Cumulative distribution of FUV-selected (dotted line), \ha-selected (dashed line), and overlapping (solid line) pixel detection fraction (DF) as a function of $R/R_{25}$. For most galaxies, \dfuv\ is higher at a given radius compared to \dfha. \dfov\ is considerably lower than \dfuv\ and \dfha\ and declines faster with radius. The vertical line shows the radii containing 90\% of the stellar mass in the galaxy, taken from \cite{padave24}. } \label{fig:detfrac}
\end{figure*}

The areal impact of FUV-selected detections is even more dramatic with galactocentric distances as seen in Figure \ref{fig:detfrac}.
Here, we show the cumulative detection fraction (DF) of the FUV-selected (\dfuv), \ha-selected (\dfha), and overlapping (\dfov) pixels as a function of $R/R_{25}$. We define DF as the ratio of pixels detected in FUV, \ha, or both to the total number of pixels detected as sources. The total number of detected pixels is estimated as $N_{UV}+N_{H\alpha}-N_{FUV+H\alpha}$, where $N_{UV}$, $N_{H\alpha}$, and $N_{FUV+H\alpha}$, are pixels detected in FUV, \ha\, and both, respectively. We subtract $N_{FUV+H\alpha}$ to avoid double counting the overlapping pixels.

Owing to the extensive areal coverage of the UV continuum distribution, we observe that \dfuv\ is generally higher at any given radius compared to the \ha-selected detections, with the exceptions of NGC 5669 and NGC 5951. This corroborates our observations in \S\ref{sec:radpro} of larger offsets between \sfruv\ and \sfrha\ seen at large galactocentric distances for the FUV-selected regions compared to the \ha-selected regions. The radial decline in \dfha\ and \dfov\ is pronounced, especially for the latter, when compared to \dfuv. The low \dfov\ suggests that a significant portion of the detections is unique to FUV and \ha\ emission. However, the decline in \dfha\ relative to \dfuv\ does not result from a decline in the overall number of \hii\ regions with radius. We detect comparable numbers of \ha- and FUV-selected regions even at large galactocentric distances (middle left panel of Figure \ref{fig:histogram}). Thus, the high \dfuv\ and reduction of overlap with increasing radius indicates that contributions from UV clumps lacking \hii\ regions are substantial.

The significant presence of FUV-only emission at larger galactocentric distances suggests that stochastic variations in the prevalence of massive stars, and consequently \hii\ regions, may increase with radius. We discuss this further in the next section.

\section{Discussion}\label{sec:disc}

\begin{figure*}[!ht]
  \centering
\includegraphics[trim = 0.3cm 0cm 25.3cm 0.cm, clip,scale=0.39]{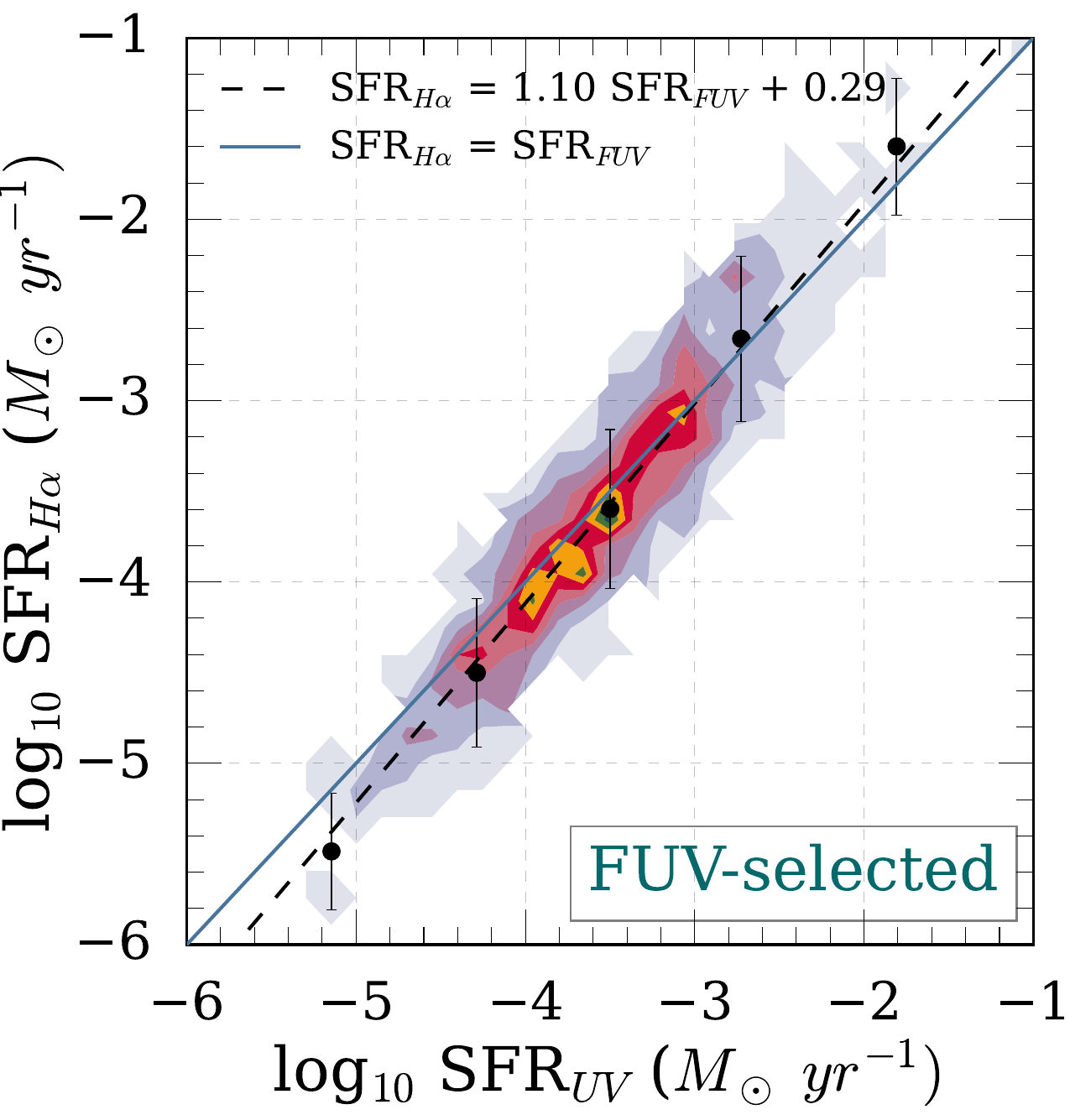}
\hfill
\includegraphics[trim = 24.5cm 0cm 0cm 0.cm, clip,scale=0.39]{figures/sfrs.pdf}
\caption{Relation between FUV and \ha\ SFRs derived from FUV-selected ({\it {left}}) and \ha-selected ({\it {right}}) regions. The solid line shows the one-to-one correspondence between the two SFRs. The dashed line shows the line of best fit. The black symbols represent median SFR values per bin with 1$\sigma$ scatter. We find a general agreement between \sfruv\ and \sfrha. } \label{fig:sfrcomp}
\end{figure*}

We investigate the relation between \sfruv\ and \sfrha\ derived from the FUV- and \ha-selected regions in Figure \ref{fig:sfrcomp}. In general, we find an agreement between the two SFRs in both cases. A simple linear regression fit to the FUV-selected data gives \sfrha\ $=1.10~(\pm~0.01)~\times$~\sfruv\ $+~0.29(\pm~0.05)$. For these regions, \sfruv\ and \sfrha\ estimates strongly correlate with Pearson-$r=0.92$. The \ha-selected estimates, on the other hand, also show a closer accord between the two indicators. We note the linear fit as \sfrha\ $=1.06~(\pm~0.01)~\times~$\sfruv\ $+~0.08~(\pm~0.04)$ with a Pearson-$r$ value of 0.93. The median SFR values per bin are depicted as black symbols and closely align with the linear trend. Additionally, we note an average scatter of $\pm$0.40 dex per SFR bin. 

\begin{figure}[!t]
  \centering
\includegraphics[trim = 0.1cm 0cm 0cm 0.cm, clip,scale=0.31]{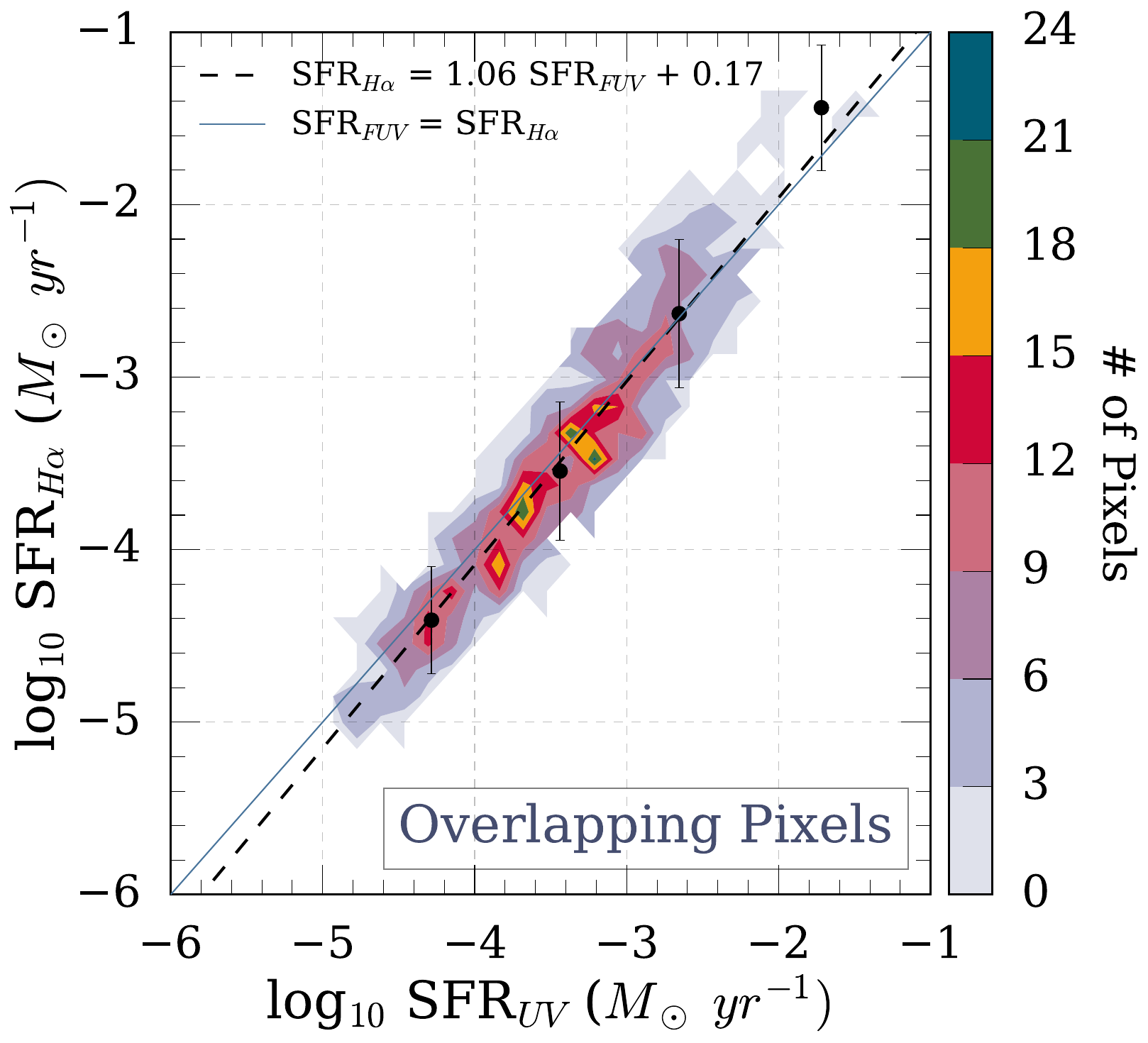}
\caption{Same as Figure \ref{fig:sfrcomp} but for overlapping pixels. } \label{fig:ovsfr}
\end{figure}

For a consistent comparison of the SFRs, we compare \sfruv\ and \sfrha\ for the overlapping pixels that are detected in both FUV and \ha\ in Figure \ref{fig:ovsfr}. The distribution of \sfruv\ and \sfrha\ for overlapping regions yields a linear fit of \sfrha\ $=1.06~(\pm~0.01)~\times$~\sfruv\ $+~0.17(\pm~0.05)$ with Pearson-$r$ of 0.94. Overall, we find consistent SFRs from the two indicators. At lower star formation levels, the underprediction of \sfrha\ is less prominent, comparable to the \ha-selected distribution.  
Furthermore, the scatter in the overall distribution is lower when compared to the FUV- and \ha-selected relations. This is expected as contributions from FUV-only and \ha-only regions are excluded in this case.

\begin{figure}[!t]
  \centering
\includegraphics[trim = 0cm 0cm 0cm 0.cm, clip,scale=0.31]{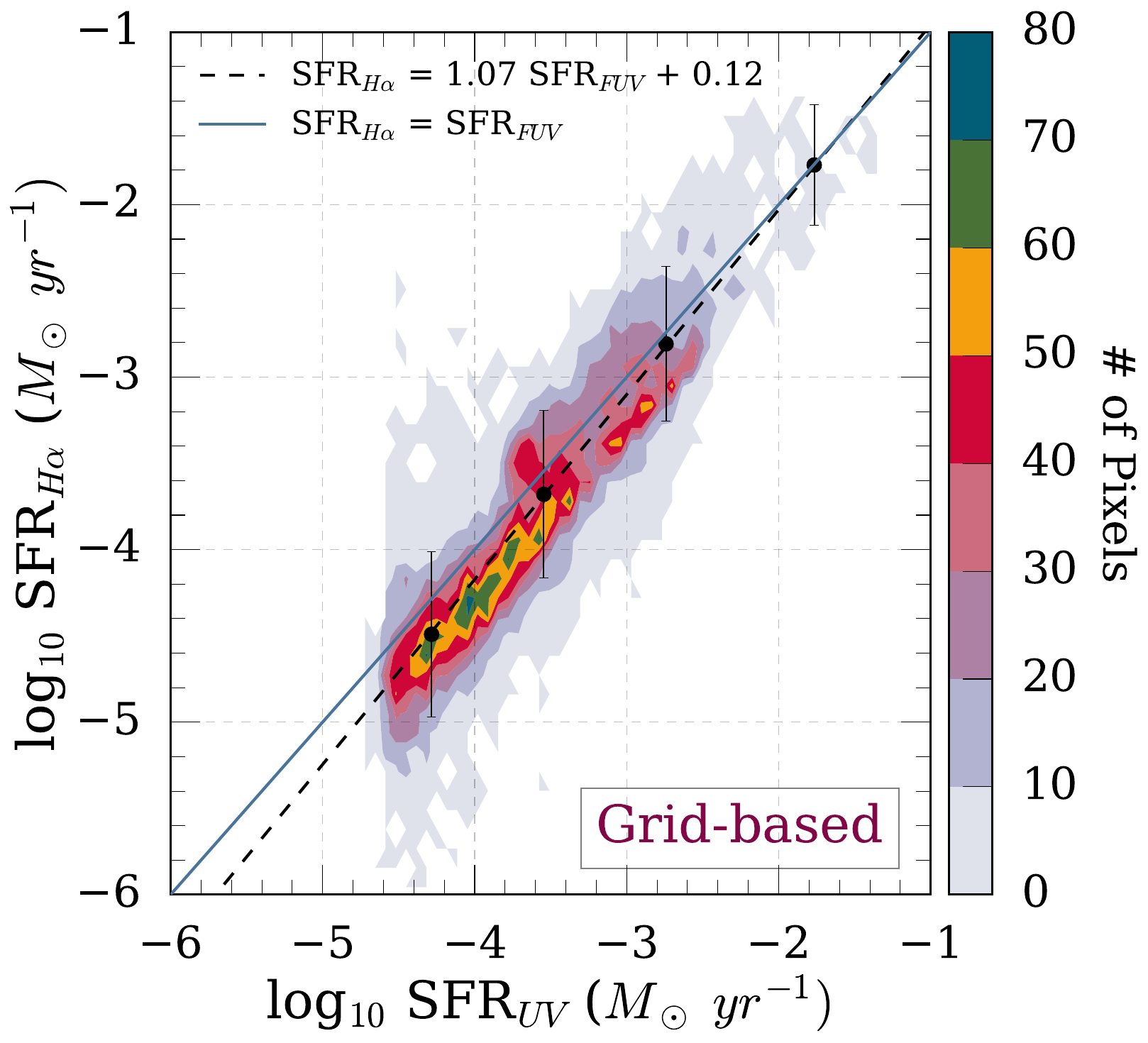}
\caption{Same as Figure \ref{fig:sfrcomp} but for Grid-based estimates.} \label{fig:grid}
\end{figure}

To completely alleviate any biases that arise from apertures, we also investigate the distribution of SFRs estimated using grid-based apertures. For this, we create 6\arcsec$\times$6\arcsec (4$\times$4 pixels) grid which is consistent with the GALEX FUV resolution. We estimate the SFRs only within grids where the FUV and \ha\ signal-to-noise ratio is above 5$\sigma$. The grid-based distribution is depicted in Figure \ref{fig:grid}. We find a larger scatter in the grid distribution, especially towards the low star formation levels. We note a linear fit giving \sfrha\ $=1.07~(\pm~0.01)~\times$~\sfruv\ $+~0.12(\pm~0.02)$ and a Pearson-$r$ of 0.87. Interestingly, we do not find the linear fit to show any significant change in comparison to the FUV- and \ha-selected fits. 

\begin{figure*}[!t]
  \centering
\includegraphics[trim = 0cm 0cm 25.4cm 0cm, clip,scale=0.39]{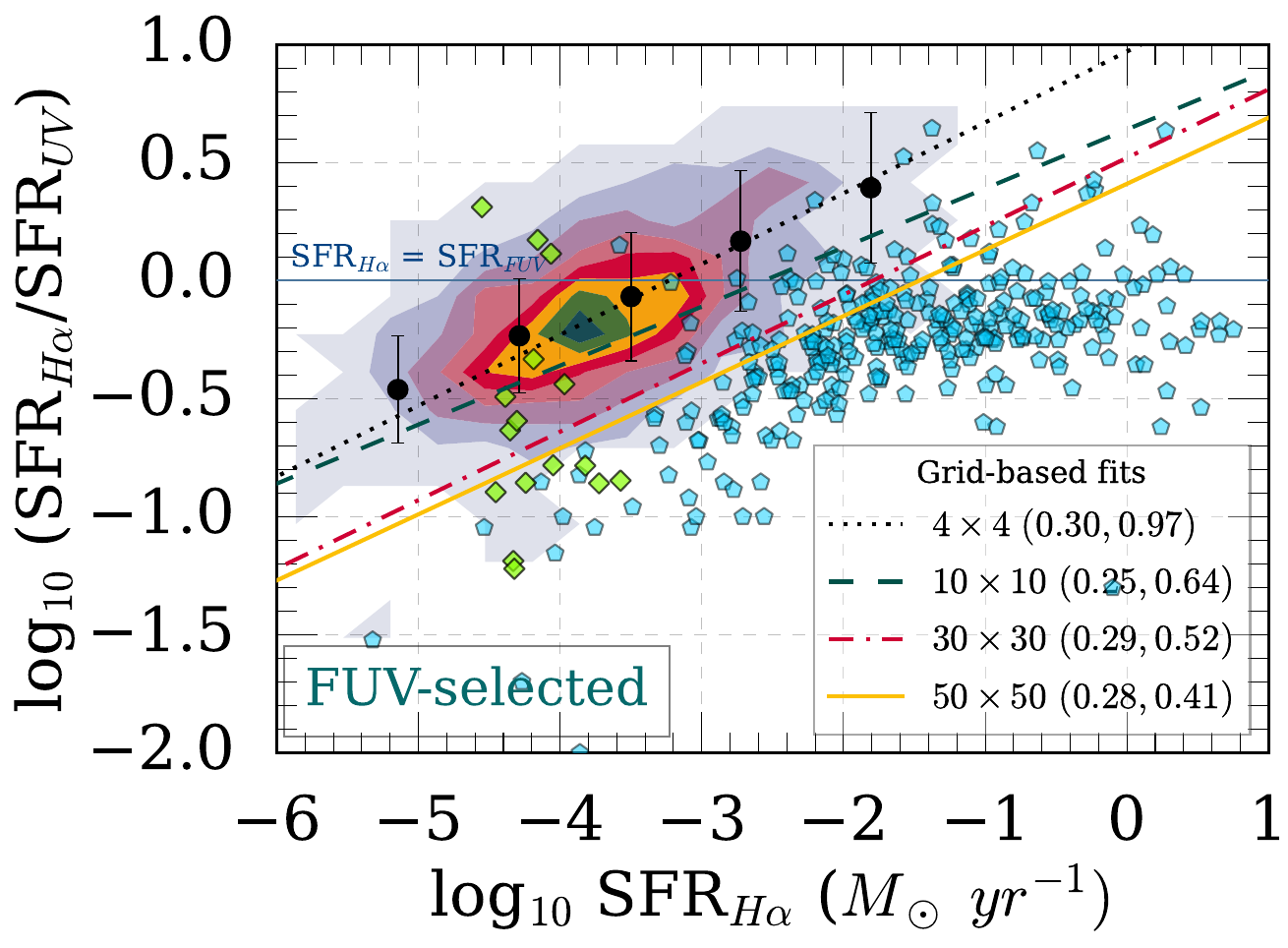}
\includegraphics[trim = 25.cm 0cm 0cm 0cm, clip,scale=0.39]{figures/hasfrs_vs_ratio.pdf}
\caption{\ha-to-FUV SFR ratios as a function of \sfrha\ for ({\it{left}}) FUV-selected and ({\it{right}}) \ha-selected regions. The solid line marks \sfruv=\sfrha. The black symbols represent median ratios and \sfrha\ per bin with 1$\sigma$ scatter. We plot integrated SFRs of star-forming spiral and dwarf galaxies from \cite{lee09} along with outlying \hii\ regions in NGC 1512/10 from \citep{werk10}. Best fit lines for grid-based SFRs are plotted in the left panel. The grid sizes, along with the parameters of the linear fit—slope and intercept—are noted on the plot. } \label{fig:ratio}
\end{figure*}

\subsection{Factors Influencing \ha-to-FUV SFR Ratios}

Our results so far have highlighted the effect of FUV- and \ha-selected apertures on \ha-to-FUV SFR ratios of star-forming regions. While in general, FUV and \ha\ closely follow each other, discrepancies in the FUV-to-\ha\ ratios seem to mainly arise from UV emission contributing more to star formation due to the inherent nature of their continuum distribution and production mechanisms. We discuss the different factors affecting the \ha-to-FUV ratios in this section.

We present the \ha-to-FUV SFR ratios of the FUV- and \ha-selected detections as a function of \sfrha\ in Figure \ref{fig:ratio}. For comparison, we include the integrated estimates for the galaxies investigated in \cite{lee09} and outlying HII regions, dubbed Emission-Line Dots (ELdots), in NGC 1512/10 from \citep{werk10}. We note that in each study, the apertures under consideration are different. While \cite{lee09}'s are integrated SFRs, \cite{werk10} use measurements from spectroscopic observations of \hii\ regions. 

We do not observe a severe decline in the ratios as is seen for integrated SFRs from \cite{lee09} at low star formation levels. This offset between our observations and the integrated values is not surprising as the latter encompasses the diffuse component of the UV emission in galaxies. We already note that UV clumps contribute significantly to star formation compared to \ha, however, the UV light associated with star clusters or clumps constitutes $\sim$20\% of the total UV flux, while the rest is dominated by diffuse light \citep{meurer95, chandar05}. This diffuse component of UV light can arise from B stars that have dispersed throughout the galaxy over time, isolated massive stars, light scattered off of dust grains, or unresolved star clusters. For nearby galaxies (z$\sim$0.5), the clumped fraction of the UV was found to constitute 15--40\% of the total UV light \citep{guo15}. For our sample, the FUV-selected regions contribute to only 8--36\% of the total \sfruv, leading to relatively high \ha-to-FUV SFR ratios when compared to integrated SFRs. Further, \ha-selected regions contribute 10--50\%, and the clump fraction is on average $\sim$1.6$\times$ higher than the UV clump fraction. 

To investigate the effect of contributions from the diffuse UV light to the \ha-to-FUV ratios, we create grids of 10$\times$10 (15\arcsec$\times$15\arcsec), 30$\times$30 (45\arcsec$\times$45\arcsec), and 50$\times$50 (75\arcsec$\times$75\arcsec) pixels, in addition to the 4$\times$4 pixel grid. The best-fit lines for the relation between \ha-to-FUV SFR ratios and \sfrha\ for the grids are plotted in the left panel of Figure \ref{fig:ratio}. 
The slopes and intercepts of the linear fits are noted in parentheses for each fit on the plot. The 4$\times$4 grid distribution aligns with the median ratios for the FUV-selected regions. However, with increasing grid sizes, the intercept is observed to decrease without much change in the slope of the distribution. These apertures capture diffuse UV flux, giving rise to \ha-to-FUV ratios $\sim$0.56 dex lower than those seen for FUV-selected regions, clearly illustrating the cause of offset between our distribution and the one from \cite{lee09}.

Larger apertures lead to lower \ha-to-FUV ratios. This is also seen when comparing the FUV-selected and \ha-selected distribution of \ha-to-FUV SFR ratios in Figure \ref{fig:ratio}. 
The distribution for our sample progressively declines for the FUV-selected regions, and tails for the \ha-selected estimates.  
At SFRs below 10$^{-4}$~M$_\odot$~yr$^{-1}$, the \ha-selected regions, \sfrha\ are generally lower than \sfruv\ by a factor of $\sim$1.2--1.5 on average. However, the FUV-selected shows that SFR from \ha\ is underpredicted by a factor of $\sim$2--3, and the offset between the two SFRs can increase to 1 dex for larger grids and integrated fluxes. 

Another factor affecting the \ha-to-FUV SFR ratios is the contribution from UV-only emission that dominates in galaxy disks (Figure \ref{fig:detfrac}), especially in the outer regions where star formation levels are low and hint at stochastic effects playing a significant role. 
In low-SFR environments, the IMF may not be fully populated leading to a lack of massive stars capable of generating \hii\ regions \citep{koda12, fuma11}.  
Combined with the FUV's ability to probe a wider range of stellar types and environments, stochastic variations in the upper end of the IMF result in an increased prevalence of UV clumps without corresponding \hii\ regions.
To illustrate stochastic effects, \cite{fuma11} used models generated from the fully stochastic stellar population synthesis code, SLUG \citep[Stochastically Lighting Up Galaxies;][]{silva12} and showed that that the \lha/\luv\ distribution in galaxies can be effectively reproduced by simultaneously modeling the SFR, cluster mass function (CMF), and IMF.  

\begin{figure}[!t]
  \centering
\includegraphics[trim = 0cm 0cm 0.cm 0.cm, clip,scale=0.28]{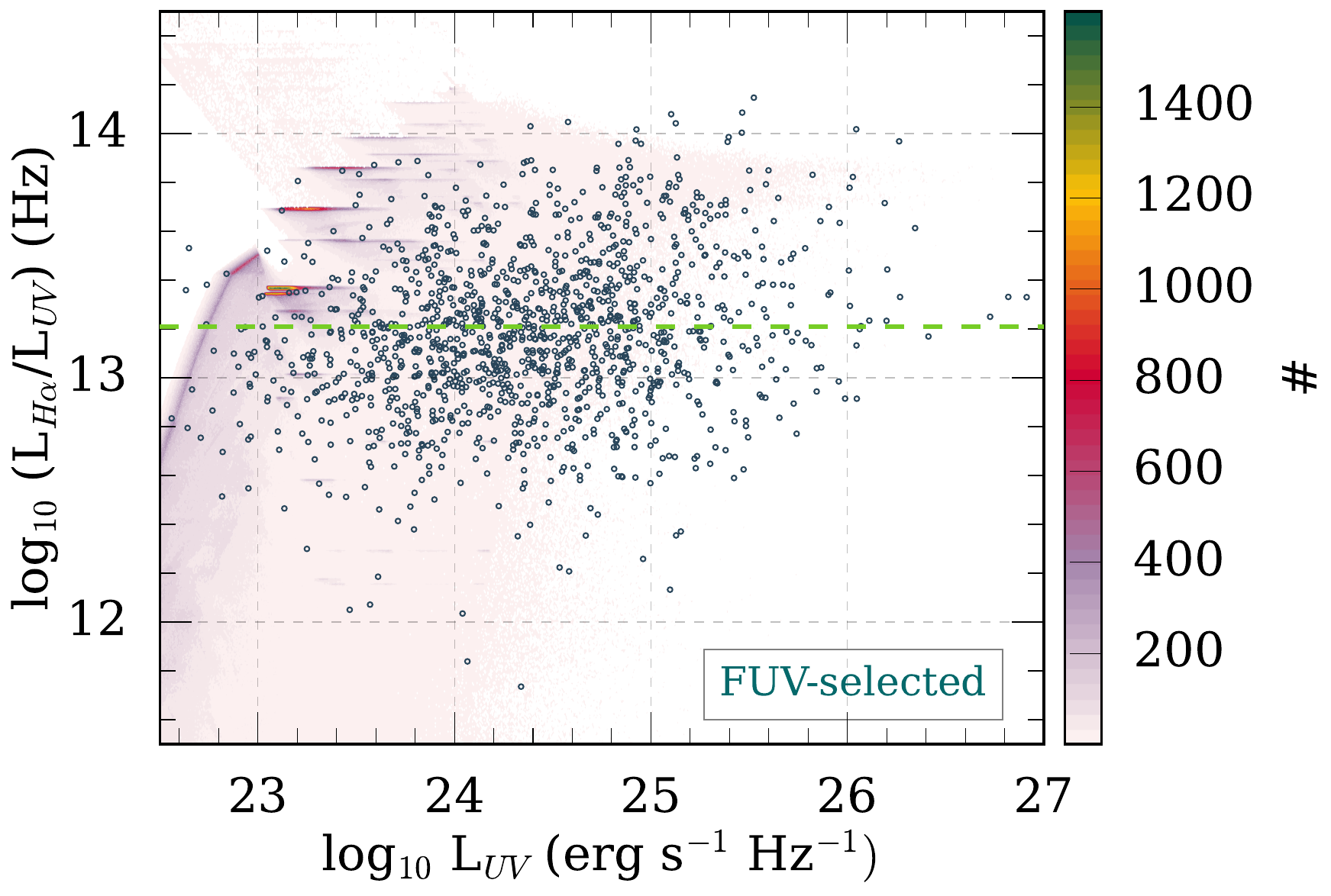}
\caption{\ha-to-FUV luminosity ratio for FUV-selected regions as a function of \luv. Models from SLUG \citep{silva12} for Kroupa IMF with a cluster fraction of 1 are plotted in the background for comparison. The dashed line represents the analytical predictions from a fully sampled Kroupa IMF. } \label{fig:slug}
\end{figure}

Drawing from \cite{fuma11}'s findings, we present the \lha/\luv\ distribution for our FUV-selected regions as a function of \luv\ in Figure \ref{fig:slug} and compare it to $\sim$5$\times$10$^{6}$ SLUG models. These models are computed for SFRs ranging from $0.2 \times 10^{-5}$ to $0.02$ M$_\odot$ yr$^{-1}$, assuming a universal Kroupa IMF spanning masses from 0.08 to 120 M$_\odot$. 
We adopt the default CMF with $\beta=2$ across the mass range of 20 to $10^7$ M$_\odot$, under the assumption that all stars form within clusters. Further details regarding the models align closely with those presented in \cite{fuma11}. Theoretical prediction from the \cite{kroupa01} IMF is also depicted in Figure \ref{fig:slug}. 
We find that our observations overlap with the simulated galaxy regions and are consistent with the analytical prediction from the IMF. While \cite{fuma11}'s investigations were carried out for integrated luminosities, with \luv$>$26~erg~s$^{-1}$~Hz$^{-1}$, we note that at sub-kpc scales the models agree with our observations.

Scatter in the \ha-to-FUV ratios also highlights other factors affecting the FUV and \ha\ emission. Flux losses in \hii\ regions are likely and result from the leakage of Lyman continuum (LyC) photons. We can infer that LyC leakage may be contributing to some extent to the underestimation of \sfrha\ from the \ha-selected and overlapping regions. While \sfrha\ and \sfruv\ are generally noted to be consistent for \ha-selected regions throughout our study, the \ha\ knots with lower \ha\ flux are likely losing ionizing photons. The leakage of ionizing photons is generally prominent in low-mass systems \citep{hoopes01, relano12} and would contribute to the diffuse component of ionized gas \citep{vanzee00, oey07}. Deeper imaging surveys like those in \cite{lee16, watkins24} and investigations of forbidden-to-Balmer line ratios \citep{hoop03, mads06, wang19} can shed light on the effect of Lyman continuum leakage and the budget of the diffuse ionizing gas on the \ha-to-FUV ratios. 

Notably, there are \hii\ regions and \ha\ knots that are found to have \sfrha$>$\sfruv. These are seen in both the FUV-selected and \ha-selected regions and throughout the galactic disk. If these regions are shrouded in dust, then FUV would be affected significantly. In that case, first-order dust corrections may not suffice. However, dust extinction effects may not be as significant in the outskirts. \cite{kimd19} investigated dust-extinction values, A$_V$ as a function of galactocentric distance for 257 nearby galaxies, and found that spiral galaxies showed steeper A$_V$ profiles. It is also likely that the outer disks have relatively small dust grains, as observed in the Milky Way and nearby galaxies \citep{zasow09, keel23}. For our sample, the level of dust emission probed using \wise\ and \mips\ data is seen to decline with radius as investigated by \cite{padave24}, suggesting that regions beyond $R_{25}$ may not be affected by dust unless these are very young and completely obscured by dust. Deeper observations in the infrared, mapping the dust and its properties would be essential \citep{will24}.

Moreover, it is also probable that regions with high \sfrha\ are going through a recent burst of star formation. FUV and \ha\ are extensively used to trace burstiness in galaxies due to the different timescales they probe \citep{emami19, mehta23, saeed23}. 
Bursty star formation histories have been extensively found to impact \ha-to-FUV SFR ratios in galaxies \citep{weisz12, emami19}. 
At the scale of star-forming regions, if a region is undergoing a starburst, the \ha\ luminosity (\lha) would increase due to contributions from new O-stars and consequently, take $\lesssim$10~Myrs to equilibrate. 
While \lha\ adjusts quickly to reflect this change in SFR, the FUV luminosity (\luv) will lag and take about $\sim$100~Myrs including contributions from the longer-lived stars. This would lead to higher \ha-to-FUV SFR ratios. Subsequently, the latter stages of the burst would result in a decline in the ratios variation in the \ha-to-FUV SFR ratios, as the FUV contribution increases and massive stars die. In lower mass systems and at low star formation levels, \cite{weisz12} found time-variable star formation histories to be consistent with their observations of \ha-to-FUV SFR ratios. 

\section{Conclusion}\label{sec:conc}

In this study, we investigate the distribution and variations in the \ha\ and FUV emission and examine \ha-to-FUV flux ratios across star-forming regions in 11 nearby star-forming galaxies. We utilize archival GALEX FUV imaging and our narrowband \ha+\nii\ imaging from the VATT to extract FUV and \ha\ sources from the respective maps. We note these sources as FUV-selected and \ha-selected regions. We investigate the demographics of these regions, morphologies, and radial distributions and also explore \sfruv\ and \sfrha\ for these regions. We summarize our key findings below: 

\begin{itemize}
    \item[1.] We detect 1335 FUV-selected and 1474 \ha-selected regions throughout our sample of 11 galaxies. The morphology of the FUV-selected regions tends to be extended and these regions generally occupy 55--85\% of the total detected area in a galaxy. \ha-selected regions, on the other hand, are generally compact compared to the FUV-selected regions. 
    \item[2.] The spatial distribution of the FUV-selected and \ha-selected regions are identical, and regions are detected even beyond $R_{25}$. However, the areal spread of FUV-selected regions is significantly larger than \ha-selected regions and increases with $R/R_{25}$. The overlap between FUV- and \ha-selected regions also decreases with $R/R_{25}$ pointing towards a radial increase in the stochastic effects in the production of massive stars responsible for producing \hii\ regions. Our observed distribution of \ha-to-FUV SFR ratios agrees with stochastic star formation models that simultaneously consider the SFR, CMF, and IMF. 
    \item[3.] Lower \ha-to-FUV SFR ratios at low levels of SFRs are primarily observed for FUV-selected regions. At SFR~$\lesssim10^{-3}$~M$_\odot$~yr$^{-1}$, \sfrha\ is underestimated by a factor of 2--3 for the FUV-selected regions. However, when larger grids are used to estimate SFRs, this factor consistently increases, and can be as high as $\sim$10. This discrepancy is due to the increasing contribution from diffuse UV flux not fully captured by FUV-selected regions, which only account for 8–36\% of the total UV flux in our sample. On the other hand, \ha-to-FUV SFR ratios for \ha-selected regions do not show any dependence on $R/R_{25}$ and \sfrha, and are only underestimated by a factor of 1.2–1.5. 
\end{itemize}

The scatter in the \sfrha\ and \sfruv\ also suggests that other factors, such as dust, LyC leakage, and varying star formation histories are also playing a role in governing the \ha-to-FUV ratio. In the future, further modeling of star formation history, metallicities, and dust effects may improve our understanding of observed offsets and how best to evaluate SFRs in low-density media.

\section*{ACKNOWLEDGEMENTS}

MP, SB, RJ, and DT are supported by NASA ADAP grant 80NSSC21K0643, SB acknowledges support from by NSF grants 2009409 and 2108159. 
MP, SB, RJ, and JM acknowledge the land and the native people that Arizona State University's campuses are located in the Salt River Valley. The ancestral territories of Indigenous peoples, including the Akimel O’odham (Pima) and Pee Posh (Maricopa) Indian Communities, whose care and keeping of these lands allow us to be here today.

We thank the staff at the Steward Observatory and the Vatican Advanced Technology Telescope for their help and support on this project. All GALEX data presented in this paper were obtained from the Mikulski Archive for Space Telescopes (MAST) at the Space Telescope Science Institute. The specific observations analyzed can be accessed via\dataset[https://doi.org/10.17909/rf6m-3y73]{https://doi.org/10.17909/rf6m-3y73}.

GALEX is a NASA Small Explorer, launched in 2003 April.
We gratefully acknowledge NASA’s support for the construction,
operation, and science analysis of the GALEX mission, developed in cooperation with the
Centre National d’´Etudes Spatiales (CNES) of France and the
Korean Ministry of Science and Technology.

Based on observations made with the NASA/ESA Hubble Space Telescope, obtained from the data archive at the Space Telescope Science Institute. STScI is operated by the Association of Universities for Research in Astronomy, Inc. under NASA contract NAS 5-26555.

This work is also partly based on observations with the VATT: the Alice P. Lennon Telescope and the Thomas J. Bannan Astrophysics Facility.

This work is based [in part] on observations made with the Spitzer Space Telescope, which was operated by the Jet Propulsion Laboratory, California Institute of Technology under a contract with NASA.

This publication makes use of data products from the Wide-field Infrared Survey Explorer, which is a joint project of the University of California, Los Angeles, and the Jet Propulsion Laboratory/California Institute of Technology, funded by the National Aeronautics and Space Administration.

This research made use of Photutils, an Astropy package for
the detection and photometry of astronomical sources \citep{larry_bradley_2023_7946442}.

\facilities{GALEX, VATT, Spitzer, WISE, VLA}

\software{astropy \citep{2013A&A...558A..33A,2018AJ....156..123A}, Photutils \citep{larry_bradley_2023_7946442}, 
          Source Extractor \citep{bert96}
         }

\bibliography{ha_ref}{}
\bibliographystyle{aasjournal}

\end{document}